%% file: main.tex
\documentclass[aps,pre,reprint,onecolumn,12pt,eprint,showkeys,raggedbottom,citeautoscript]{revtex4-2}

\usepackage[T1]{fontenc}
\usepackage[english]{babel}
\usepackage{microtype, setspace}
\usepackage{amsmath, amssymb, amsthm, amsfonts, mathtools, mathrsfs}
\usepackage{csquotes, enumitem, moreenum}
\usepackage{ifthen, comment}
\usepackage[caption=false]{subfig}
\usepackage[dvipsnames]{xcolor}
\usepackage{relsize}
\usepackage{tikz}
\usepackage{hyperref}
\usepackage{xurl}
\usepackage[capitalize]{cleveref}
\hypersetup{colorlinks=true, allcolors=blue!50!black, breaklinks=true}

\crefname{section}{Sec.}{Sects.}

\newcommand{\ce}{\coloneqq}
\newcommand{\ec}{\eqqcolon}
\newcommand{\wt}[1]{\widetilde{#1}}

\newcommand{\R}{\mathbb{R}}

\newcommand{\MD}{\mathcal{D}}
\newcommand{\MH}{\mathcal{H}}
\newcommand{\ML}{\mathcal{L}}
\newcommand{\MT}{\mathcal{T}}
\newcommand{\MO}{\mathcal{O}}
\newcommand{\ii}{\mathrm{i}}
\newcommand{\ee}{\mathrm{e}}
\newcommand{\id}{\mathop{}\!\mathrm{Id}}
\newcommand{\diff}{\mathop{}\!\mathrm{d}}

\newcommand{\bfr}{\mathbf{r}}
\newcommand{\bfp}{\mathbf{p}}
\newcommand{\set}[2][]{#1\{{#2}#1\}}

\newcommand{\ket}[1]{\mathinner{|{#1}\rangle}}
\newcommand{\bra}[1]{\mathinner{\langle{#1}|}}
\newcommand{\braket}[2][]{#1\langle{#2}#1\rangle}

\newcommand{\ketbra}[2]{\ket{#1} \! \bra{#2}}
\newcommand{\dbraket}[4][]{#1\langle{#2}\,#1\vert\,{#3}\,#1\vert\,{#4}#1\rangle}
\DeclareMathOperator{\tr}{tr}

\newtheorem*{theorem}{Theorem}
\newtheorem*{conclusion}{Conclusion}
\theoremstyle{definition}
\newtheorem*{remark}{Remark}

\usetikzlibrary{decorations.pathmorphing}
\tikzstyle{spring}=[thick, decorate, decoration={coil, amplitude=3pt, segment length=3pt}]
\tikzstyle{trajectory}=[thick, decorate, decoration={snake, amplitude=3pt, segment length=1.75cm}]
\newlength{\reflen}
\colorlet{nodeGreen}{OliveGreen}
\colorlet{intBlue}{NavyBlue}

\newcommand{\FUBaffiliation}{\affiliation{Freie Universität Berlin, Institute of Mathematics, Arnimallee 6, 14195 Berlin, Germany}}

\begin{document}

\title{Open quantum systems beyond equilibrium: Lindblad equation and path integral molecular dynamics}

\author{Benedikt M. Reible}
\email{benedikt.reible@fu-berlin.de}
\author{Somayeh Ahmadkhani}
\email{s.ahmadkhani@fu-berlin.de}
\author{Luigi Delle Site}
\email{luigi.dellesite@fu-berlin.de}
\FUBaffiliation

\begin{abstract}
    The Lindblad equation determines the time evolution of the density operator of open quantum systems. While valid for any system size, its use is, in practice, restricted to prototype/surrogate models with the aim of tackling specific aspects of the overall quantum complexity of a multi-atomic system. Path integral molecular dynamics (PIMD) instead provides static and dynamical quantum statistical averages of physical observables for systems in equilibrium composed of up to thousands of atoms over timescales up to nanoseconds, under the condition that short-time quantum coherence is not relevant for the properties of interest. PIMD relies on the well-established technique of molecular dynamics (MD) with its associated classical trajectories. However, it cannot describe a direct time evolution of a system and its convergence to a stationary state in situations out of equilibrium. In this work, we analyze the link between the Lindblad equation and PIMD; specifically, we will discuss how PIMD can actually be used to calculate the time evolution of ensemble-averaged physical observables and their convergence to a stationary state for situations out of equilibrium, bypassing the need of explicitly solving the Lindblad equation. Yet, at the same time, the Lindblad equation and PIMD are linked to one another through a formal relation of equivalence, which provides an argument for the consistency of PIMD results, namely the positivity of the density operator at any time. A numerical study of a prototype system, which is of interest in chemical physics, will be used to showcase the method.
\end{abstract}

\maketitle

\section{Introduction}

Open quantum systems \cite{BreuerPetruccione2002, RivasHuelga2012, Vacchini2024} provide, at the very foundational level, a theoretical background for the study of the physics related to the rapidly developing quantum materials. The pursuit of energy-efficient systems \cite{Goyal2025}, advances in quantum information \cite{Xavier2025}, and even the long-term vision of practical quantum technologies and computers \cite{Acin2018} all rely on the design of such modern materials. The research pursuing these goals provides fertile ground, where abstract concepts lay the theoretical foundation \cite{Prior2010, McCauley2020} and prepare the ground for generic physical models that test these foundations and yield fundamental predictions \cite{Braaten2017, Guimaraes2016, Bose2024, Reible2025apl} which, in turn, must be made concrete by considering specific materials with chemical accuracy at the level of single atoms and molecules \cite{Morais2023, Singh2023, Chattaraj2025}. Such a synergetic process allows for the design of quantum materials with properties on demand that can fulfill any of the aforementioned requirements.

The cross-fertilization of fields and disciplines promoted by this kind of research is also a characteristic of the present work: on one hand, we discuss the celebrated Lindblad equation \cite{Lindblad1976, Gorini1976} (see also Refs. \cite{BreuerPetruccione2002, RivasHuelga2012, Vacchini2024, Manzano2020}), which allows to examine fundamental properties of a quantum system at the level of single elementary quantum components (e.g., spin, photon, phonon, qubit) in and out of equilibrium, treated at time scales that still capture their quantum coherence and decoherence. The Lindblad equation becomes impractical, however, for many-atom systems, where the quantum properties of interest involve a specific chemical design. In this perspective, we discuss, on the other hand, molecular dynamics (MD) in its quantum version using Feynman's path integral \cite{Feynman1948}, termed path integral molecular dynamics (PIMD); see, for example, Ref. \cite{Tuckerman2023}. The PIMD method can handle thousands of atoms with their specific chemistry at the cost of losing the capability to describe the elementary physical degrees of freedom mentioned above and being restricted to treating time scales that are certainly large enough for mesoscopic properties, but not small enough to capture quantum coherence. Nevertheless, many condensed matter systems and corresponding processes of interest are characterized by rapid decoherence, and the corresponding level of \enquote{quantumness} captured by PIMD is sufficient for designing atomistic systems with specific mesoscopic characteristics needed in modern quantum technology. A representative example is the heat transport in carbon nanotubes \cite{Liang2025} that may be optimized by specifically designing its atomistic structure \cite{Donadio2007, Barbalinardo2021, Benenti2023}. In this case, the phononic transport described in terms of the Lindblad equation would be possible only for a highly simplified surrogate system with results that are not directly applicable to realistic systems with chemical accuracy. Instead, the MD approach can directly deliver a suggestion for the specific chemistry that leads to the optimal characteristics required for the desired properties (see, e.g., Ref. \cite{Donadio2007}).

The impression one might get from the above discussion is that the two approaches are quite separated and actually share little from the conceptual and practical point of view; however, this is actually not the case. Observe that the PIMD method has been restricted to systems in equilibrium so far, where the form of the density matrix is known \textit{a priori} so that sampling of the latter is well-posed and can be checked \textit{a posteriori}. The technique has not been proposed for situations out of equilibrium precisely because in such a case, the density matrix is not known and PIMD can only implicitly sample it; hence, one needs criteria that assure positivity of the density matrix at any given time. It is well-known that the Lindblad equation assures positivity of the density matrix during the entire time evolution for systems in and out of equilibrium. Based on these observations, we formalize a framework to extend the PIMD idea to situations out of equilibrium and discuss arguments that assure its internal quantum-mechanical and statistical-mechanical consistency. At the same time, we conclude that the Lindblad equation provides a mandatory criterion for the positivity of the corresponding density matrix in the extended PIMD framework.

As an example for an application of the method, we will treat the case of a one-dimensional chain of water molecules in a thermal gradient. The steady state of a uniform chain under the effect of a temperature gradient is well-known; thus, if we apply our method, and if the method is correct, we should reproduce the known behavior. Moreover, the quantum effect of the path integral representation must be consistent with the expected physical behavior of atoms delocalized in space under the effect of a heat flow. We will show that our approach indeed gives physically consistent and numerically satisfactory results. The reason for choosing a one-dimensional water chain as a showcase is threefold: (i) we need a system that is simple enough, from the physical point of view, so that the results may have a clear interpretation. Yet, (ii) the system should be technically challenging for testing the numerical solidity of the approach. Of course, (iii) the system should also be complex enough regarding its chemical composition so that one can see the practical utility of our approach for systems of interest that cannot be treated via the standard Lindblad approach. Regarding the last point, it must be noted that one-dimensional water chains are a prototype of water wires, that is, a typology of system that has gained attention in a large field of chemical physics \cite{jctc,davide-julia,pnas}. Once again, we underline that for a system as simple as the one treated by us here, it would not be feasible to solve the Lindblad equation with standard computational resources. Certainly, if one could use the latter, the problem would be solved at the full electronic and nuclear scale; however, if one wants to study, for example, the quantum effect on heat transport, then our method can deliver physically sound results without the need of large computational resources.

The paper is structured as follows: In the subsequent \cref{sec:LindbladPIMD}, we will discuss the Lindblad equation in connection with the PIMD method and elaborate on how we intend to connect the two. \cref{sec:PI} contains a brief review of the derivation of the path integral representation of the canonical density matrix, which is used in \cref{sec:PIMD} to introduce the PIMD method for computing quantum-mechanical ensemble averages. Our main proposal is contained in \cref{sec:beyondEquilibrium}, where we will discuss an extension of the classical D-NEMD method to quantum systems and how this method, together with the PIMD formalism, can be used to compute averages of non-equilibrium observables. Finally, in \cref{sec:application} we will report the results of the method applied to a system of chemical physics.

\section{From the Lindblad equation to PIMD}\label{sec:LindbladPIMD}

The Lindblad equation is a quantum master equation governing the time evolution of the density matrix of an open quantum system. It is derived from the von Neumann equation by dividing the total system into a subsystem of interest (the \enquote{open system}) and a much larger environment (the \enquote{reservoir}) whose degrees of freedom are traced out, see, e.g., Refs. \cite{BreuerPetruccione2002, RivasHuelga2012, Vacchini2024, Manzano2020} and Appendix \ref{app:Lindblad} for a short overview of the derivation. The final equation takes the following form:
\begin{equation}\label{eq:Lindblad}
    \frac{\diff \hat{\rho}(t)}{\diff t} = -\frac{\ii}{\hbar} \, [\hat{H}, \hat{\rho}] + \sum_{j} \lambda_j \left(\hat{L}_j \hat{\rho}(t) \hat{L}_j^\dagger - \frac{1}{2} \hat{L}_j^\dagger \hat{L}_j \hat{\rho}(t) - \frac{1}{2} \hat{\rho}(t) \hat{L}_j^\dagger \hat{L}_j\right) \ .
\end{equation}
Here, $\hat{\rho}$ is the density operator and $\hat{H}$ the Hamiltonian of the open system, the operators $\hat{L}_j$ describe the dissipative interaction with the reservoir, and the constants $\lambda_j \ge 0$ are the damping rates.

Calculating the time dependence $t \mapsto \hat{\rho}(t)$ using the Lindblad equation is an essential task because it facilitates the computation of time-dependent statistical averages of any physical observable $\hat{A}$ of the open system:
\begin{equation}\label{eq:timeDependentExpectation}
    \braket{\hat{A}}(t) \ce \tr \bigl(\hat{\rho}(t) \hat{A}\bigr) \ .
\end{equation}
Moreover, knowing the function $t \mapsto \hat{\rho}(t)$ allows drawing explicit conclusions about the convergence of the system to a stationary state of equilibrium $\hat{\rho}_\mathrm{eq}$ or non-equilibrium $\hat{\rho}_\mathrm{neq}$ in the limit $t \to + \infty$; typical cases include the evolution of the system to a stationary state of equilibrium through thermal equilibration (i.e., the system is in contact with a thermostat), systems characterized by the presence of a heat flux as examples of non-equilibrium situations (i.e., the system is in contact with two distinct heat baths), or generally a response of the system to any external perturbation. A crucial point for our work is that whenever the coupling of the open system with the environment can be written in the form of \cref{eq:Lindblad}, one can be assured that for all times $t$, the density matrix $\hat{\rho}(t)$ is a positive operator (see Lindblad's original paper \cite{Lindblad1976} as well as \cite[Theorem 4.2.1]{RivasHuelga2012}, \cite[Theorem 5.1]{Vacchini2024}).

For systems composed of hundred of atoms, solving the Lindblad equation is prohibitive, hence the question arises whether there are techniques with the help of which one can determine the density matrix at an acceptable computational price and if so, how they are related to the Lindblad equation. One interesting example of a technique that numerically samples the density matrix for systems in equilibrium is the path integral molecular dynamics (PIMD) method \cite{Tuckerman2023, Jang1999, Habershon2013}. By mapping the quantum system onto a classical system, this technique allows for the implicit sampling of the equilibrium density matrix through classical trajectories; that is, quantum statistical averages of physical observables (in equilibrium) are calculated by averaging the observable of interest over the  trajectories of the surrogate classical system. However, in comparison to the Lindblad approach which automatically handles situations beyond equilibrium, applying the PIMD method to systems out of equilibrium is not straightforward. One major result of the present work is that we analyze the problem of going beyond equilibrium and, using the theoretical structure of out-of-equilibrium techniques known from classical MD, put forward a model which describes, consistently with the quantum formalism, situations out of equilibrium in the PIMD formulation. Specifically, we focus on how to calculate expectation values of the form \eqref{eq:timeDependentExpectation} for general physical observables $\hat{A}$ via PIMD, subject to the condition that
\begin{equation*}
    \lim_{t \to + \infty} \braket{\hat{A}} (t) = \tr \bigl(\hat{\rho}_\mathrm{neq} \hat{A}\bigr) \ec \braket{\hat{A}}_\mathrm{neq} \ .
\end{equation*}

Through our analysis, we find a very important connection between the Lindblad equation and PIMD applied to situations out of equilibrium, viz., the former dictates a necessary condition that the latter must satisfy in order to assure physical consistency of the time-dependent expectation values \eqref{eq:timeDependentExpectation}. Such a condition imposes that in the PIMD simulation setup, the coupling of the system with any external source must be compatible with (or mappable onto) the dissipative term in the Lindblad equation \eqref{eq:Lindblad}:
\begin{equation}\label{eq:LindbladDissipative}
    \MD \bigl(\hat{\rho}(t)\bigr) = \sum_{j} \lambda_j \left(\hat{L}_j \hat{\rho}(t) \hat{L}_j^\dagger - \frac{1}{2} \hat{L}_j^\dagger \hat{L}_j \hat{\rho}(t) - \frac{1}{2} \hat{\rho}(t) \hat{L}_j^\dagger \hat{L}_j\right) \ .
\end{equation}
This condition assures that the corresponding $\hat{\rho}(t)$, implicitly sampled by the PIMD method in situations out of equilibrium, is positive for any time $t$.

\section{Path integral representation of the density matrix}\label{sec:PI}

In this section, we briefly outline the derivation of the well-known path integral representation of the canonical density matrix for a system of particles in equilibrium, which can be found, for example, in Refs. \cite[Chapter 12]{Tuckerman2023} and \cite[Chapter 10]{FeynmanHibbs2010} (we follow the former closely; for mathematical aspects of the path integral see, e.g., Ref. \cite{Mazzucchi2022}).

Consider first a single quantum particle moving in one-dimensional space; once the formula is derived for this case, the extension to a many-particle system in three-dimensional space is straightforward. Let the Hamiltonian of the particle be given by
\begin{equation*}
    \hat{H} = \frac{\hat{p}^2}{2m} + \hat{U}(\hat{x}) \equiv \hat{K} + \hat{U} \ ,
\end{equation*}
where $\hat{p} = - \ii \hbar \partial_x$ is the momentum operator, $\hat{x}$ the position operator, $\hat{K} = \hat{p}^2 / 2m$ the kinetic energy operator, and $\hat{U}$ the potential energy operator. For all $x, x^\prime \in \R$ define
\begin{equation*}
    \rho(x, x^\prime; \beta) \ce \braket[\big]{x^\prime \,\big\vert\, \ee^{- \beta \hat{H}} \,\big\vert\, x} \ ,
\end{equation*}
which may be interpreted as the position-space matrix elements of the unnormalized trace-class operator $\hat{\rho} = \ee^{- \beta \hat{H}}$. Since $\hat{K}$ and $\hat{U}$ do not commute in general, one has to resort to an approximation theorem to factorize the exponential $\ee^{- \beta \hat{H}}$, namely the Trotter product formula \cite{Trotter1959, Nelson1964, Chernoff1968} (see also Ref. \cite[Section VIII.8]{RS1}):
\begin{equation*}
    \rho(x, x^\prime; \beta) = \lim_{P \to +\infty} \braket[\Big]{x^\prime \,\Big\vert\,\mathopen{} \bigl(\underbrace{\strut\ee^{- \beta \hat{U} / (2 P)} \, \ee^{- \beta \hat{K} / P} \, \ee^{- \beta \hat{U} / (2 P)}}_{\ec \hat{W}}\bigr)^P \,\Big\vert\,\mathopen{} x} = \lim_{P \to +\infty} \braket[\bigg]{x^\prime \,\bigg\vert\,\mathopen{} \prod_{j=1}^{P} \hat{W} \,\bigg\vert\,\mathopen{} x} \ .
\end{equation*}
Inserting $(P-1)$-times the position-space spectral representation of the identity operator, $\id = \int_\R \diff x_j \ketbra{x_j}{x_j}$, $j \in \set{2, \dotsc, P}$, in between the factors of $\hat{W}$ and simplifying the resulting expectation values using some standard arguments, it follows that \cite[pp. 492 ff.]{Tuckerman2023}
\begin{align*}
    \rho(x, x^\prime; \beta) &= \lim_{P \to +\infty} \left(\frac{m P}{2 \pi \beta \hbar^2}\right)^{P/2} \int_\R \dotsi \int_\R \diff x_2 \dotsm \diff x_P \\
    &\qquad \times \exp \left.\left\{- \frac{1}{\hbar} \sum_{j=1}^{P} \left[\frac{m P}{2 \beta \hbar} \, (x_{j+1} - x_j)^{2} + \frac{\beta \hbar}{2 P} \, \bigl(U(x_{j+1}) + U(x_j)\bigr) \right] \right\} \right|_{x_1 = x}^{x_{P+1} = x^\prime} \ ,
\end{align*}
where $U(x_j)$ denotes the eigenvalue of the operator $\hat{U} (\hat{x})$ corresponding to the generalized position eigenvector $\ket{x_j}$, $j \in \set{1, \dotsc, P + 1}$.

Suppose now that the particle is confined to the spatial region $[0, L]$ for some $L > 0$. Then one may compute the canonical partition function $Z = \tr(\ee^{- \beta \hat{H}})$ by evaluating the trace in the basis of generalized eigenvectors $\ket{x}$:
\begin{equation*}
    Z = \tr(\ee^{-\beta \hat{H}}) = \int_{0}^{L} \braket{x \,\vert\, \ee^{- \beta \hat{H}} \,\vert\, x} \diff x = \int_{0}^{L} \rho(x, x; \beta) \diff x \ .
\end{equation*}
Setting $x_1 = x_{P+1} = x$ in the expression for $\rho(x, x^\prime; \beta)$ from above and using that in this case, $\sum_{j=1}^{P} \bigl(U(x_{j+1}) + U(x_j)\bigr) = 2 \sum_{j=1}^{P} U(x_j)$, it follows that
\begin{align}\label{eq:partitionFunctionPI}
    \begin{split}
        Z &= \lim_{P \to +\infty} \left(\frac{m P}{2 \pi \beta \hbar^2}\right)^{P/2} \int_{0}^{L} \dotsi \int_{0}^{L} \diff x_1 \dotsm \diff x_P \\
        &\qquad \times \exp \left.\left\{- \frac{1}{\hbar} \sum_{j=1}^{P} \left[\frac{m P}{2 \beta \hbar} \, (x_{j+1} - x_j)^2 + \frac{\beta \hbar}{P} \, U(x_j) \right] \right\} \right|_{x_{P+1} = x_1} \ ,
    \end{split}
\end{align}
where the integration variable $x$ has been renamed to $x_1$; this formula is the \emph{discretized path integral representation} of the partition function of a single particle. Observe the following crucial fact: defining the \enquote{effective} classical Hamiltonian function
\begin{equation*}
    H_P (x_1, \dotsc, x_P) \ce \sum_{j=1}^{P} \left[\frac{1}{2} \, m \omega_P^2 (x_j - x_{j+1})^{2} + \frac{1}{P} \, U(x_j) \right]
\end{equation*}
with $\omega_P = \sqrt{P} / (\beta \hbar)$ and $x_{P+1} = x_1$, it follows that the partition function can be written more compactly as
\begin{equation*}
    Z = \lim_{P \to +\infty} \left(\frac{m P}{2 \pi \beta \hbar^2}\right)^{P/2} \int_{0}^{L} \dotsi \int_{0}^{L} \ee^{- \beta H_P (x_1, \dotsc, x_P)} \diff x_1 \dotsm \diff x_P \ .
\end{equation*}
This expression can be seen as the partition function of a polymer ring, that is, a classical object with $P$ beads harmonically linked to each other via nearest-neighbor connections, with $\omega_P$ being the coupling strength (see \cref{fig:pimd} below). This observation is the starting point for the PIMD method discussed in the next section.

Before we come to this, we mention how the above formulas can be extended to $N$-particle systems in three-dimensional space. This extension is, in principle, complicated because the bosonic/fermionic symmetry of the particles must be respected \cite[Sec. 12.6]{Tuckerman2023}. However, in first approximation, basic quantum effects of particle delocalization do not require the introduction of exchange symmetry. Therefore, it follows that for an $N$-particle Hamiltonian of the form $H = \sum_{i=1}^{N} \hat{\bfp}_i^2 / (2 m_i) + \hat{U} (\hat{\bfr}_1, \dotsc, \hat{\bfr}_N)$, one has \cite[Eq. (12.6.11)]{Tuckerman2023}:
\begin{align}\label{eq:partitionFunctionPolymer}
    \begin{split}
        Z &= \lim_{P \to +\infty} \, \left[\,\prod_{i=1}^{N} \left(\frac{m_i P}{2 \pi \beta \hbar^2}\right)^{3P/2}\right] \int_{D(L)} \, \prod_{i=1}^{N} \diff^3 \bfr_i^{(1)} \dotsm \diff^3 \bfr_i^{(P)} \\
        &\qquad \times \exp \left.\left\{- \sum_{j=1}^{P} \left[\sum_{i=1}^{N} \frac{m_i P}{2 \beta \hbar^2} \, \bigl(\bfr_i^{(j+1)} - \bfr_i^{(j)}\bigr)^2 + \frac{\beta}{P} \, U \bigl(\bfr_1^{(j)}, \dotsc, \bfr_N^{(j)}\bigr) \right] \right\}\right|_{\bfr_i^{(P+1)} = \bfr_i^{(1)}} \ .
    \end{split}
\end{align}
Here, the integration region $D(L)$ is a $3NP$-dimensional hypercube of side length $L > 0$, and $U \bigl(\bfr_1^{(j)}, \dotsc, \bfr_N^{(j)}\bigr)$ denotes the eigenvalue of $\hat{U} (\hat{\bfr}_1, \dotsc, \hat{\bfr}_N)$ corresponding to $\ket{\bfr_1^{(j)}, \dotsc, \bfr_N^{(j)}}$. It is important to note that, according to this formula, the path integral formalism imposes that in the polymer-ring representation, the inter-particle interaction $\hat{U}$ is realized only in terms of an interaction between beads with the same index $j$ for each individual particle.

By virtue of \cref{eq:partitionFunctionPolymer}, an $N$-particle quantum problem has been mapped to a classical problem of interacting polymer rings, ideal for classical MD. In fact, using an effective fictitious dynamics of polymer rings, classical MD allows for the calculation of quantum statistical averages of physical observables, as reported below.

\section{Path Integral Molecular Dynamics}\label{sec:PIMD}

It is well-known that classical MD samples the probability distribution of classical systems in phase space, thus allowing for the computation of ensemble averages of physical observables by averaging their value over MD trajectories \cite{Tuckerman2023}. In the present context, the probability distribution in phase space of polymer rings effectively represents the quantum density matrix, as shown above, and hence one may try to compute quantum ensemble averages of physical observables by averaging their classical counterpart, written in the polymer-ring formalism, over the classical trajectories of the polymer-ring  system. To realize this method, an additional step is needed since MD simulations require momenta associated to each particle, which are not present in the effective polymer-ring Hamiltonian, cf. \cref{eq:partitionFunctionPolymer}.

The problem can be solved by artificially adding $P$ Gaussian integrals over fictitious momentum variables $p_1, \dotsc, p_P$ with mass parameter $m^\prime = m P / (2 \pi \hbar)^2$ \cite[p. 525]{Tuckerman2023}. Indeed, for a single particle we may rewrite the prefactor in \cref{eq:partitionFunctionPI} as
\begin{equation*}
    \left(\frac{m P}{2 \pi \beta \hbar^{2}}\right)^{P/2} = \int_\R \dotsi \int_\R \exp \left(- \beta \sum_{j=1}^{P} \frac{p_j^2}{2 m^\prime}\right) \diff p_1 \dotsm \diff p_P \ .
\end{equation*}
For the $N$-particle system, we assume, for simplicity, that the particles have identical mass: $m_i \equiv m$ for all $i \in \set{1, \dotsc, N}$. The partition function \eqref{eq:partitionFunctionPolymer} then takes the following form:
\begin{align*}
    &Z = \lim_{P \to +\infty} \, \int_\R \, \prod_{i=1}^{N} \diff^3 \bfp_i^{(1)} \dotsm \diff^3 \bfp_i^{(P)} \int_{D(L)} \, \prod_{i=1}^{N} \diff^3 \bfr_i^{(1)} \dotsm \diff^3 \bfr_i^{(P)} \\
    &\times \exp \left.\left\{- \sum_{j=1}^{P} \left[\sum_{i=1}^{N} \frac{\beta}{2 m^\prime} \, \bigl(\bfp_i^{(j)}\bigr)^2 + \frac{m P}{2 \beta \hbar^2} \, \bigl(\bfr_i^{(j+1)} - \bfr_i^{(j)}\bigr)^2 + \frac{\beta}{P} \, U \bigl(\bfr_1^{(j)}, \dotsc, \bfr_N^{(j)}\bigr) \right] \right\}\right|_{\bfr_i^{(P+1)} = \bfr_i^{(1)}} \ .
\end{align*}
Thus, the effective classical polymer-ring Hamiltonian is
\begin{align}\label{eq:effectiveQuantumH}
    \begin{split}
        H_P \bigl(\bigl\{\bfr_i^{(j)}\bigr\}, \bigl\{\bfp_i^{(j)}\bigr\}\bigr) \ce \sum_{j=1}^{P} \Biggl\{ &\sum_{i=1}^{N} \left[ \frac{1}{2 m^\prime} \, \bigl(\bfp_i^{(j)}\bigr)^2 + \frac{1}{2} \, m \omega_P^2 \bigl(\bfr_i^{(j+1)} - \bfr_i^{(j)}\bigr)^2 \right] \\
            & + \frac{1}{P} \, U \bigl(\bfr_1^{(j)}, \dotsc, \bfr_N^{(j)}\bigr) \Biggr\}
    \end{split}
\end{align}
with $\omega_P = \sqrt{P} / (\beta \hbar)$ and $\bfr_i^{(P+1)} = \bfr_i^{(1)}$ as before, and the curly braces in the argument of $H_P$ indicate that this function depends on all $\bfr_i^{(j)}, \bfp_i^{(j)}$, $i \in \set{1, \dotsc, N}$, $j \in \set{1, \dotsc, P}$. The partition function now reduces to
\begin{equation*}
    Z = \lim_{P \to +\infty} \, \int_\R \, \prod_{i=1}^{N} \diff^3 \bfp_i^{(1)} \dotsm \diff^3 \bfp_i^{(P)} \int_{D(L)} \, \prod_{i=1}^{N} \diff^3 \bfr_i^{(1)} \dotsm \diff^3 \bfr_i^{(P)} \, \exp \Bigl(- \beta H_P \bigl( \bigl\{\bfr_i^{(j)}\bigr\}, \bigl\{\bfp_i^{(j)}\bigr\} \bigr) \Bigr) \ .
\end{equation*}
Let $Z_P$ denote the $6NP$-dimensional integral over $\ee^{- \beta H_P}$, i.e., the terms after the limit $P \to + \infty$ in the above expression. We then have $Z = \lim_{P \to + \infty} Z_P$, implying that the quantity $Z_P$ may be viewed as a classical approximation of accuracy $P$ for the exact quantum-mechanical partition function $Z$. The number of beads $P$ controls the accuracy since the larger this number, the closer the classical expression $Z_P$ will be to the quantum-mechanical $Z$ (see also \cref{subsec:accuracy} below).

Consider a physical observable $\hat{A}$ which is a function of the position operators $\hat{\bfr}_1, \dotsc, \hat{\bfr}_N$ only, and let $A(\bfr_1, \dotsc, \bfr_N)$ denote its eigenvalue corresponding to $\ket{\bfr_1, \dotsc, \bfr_N}$. In order to treat the $P$ different beads $\bfr_i^{(j)}$, $j \in \set{1, \dotsc, P}$, for each particle $i$ in the polymer-ring formalism equally, one can introduce the following estimator \cite[pp. 498 f.]{Tuckerman2023}:
\begin{equation}\label{eq:classicalEstimator}
    A_P \bigl( \bigl\{\bfr_i^{(j)}\bigr\} \bigr) \ce \frac{1}{P} \sum_{j=1}^{P} A \bigl( \bfr_1^{(j)}, \dotsc, \bfr_N^{(j)} \bigr) \ .
\end{equation}
Within the polymer-ring formalism, the equilibrium average $\braket{\hat{A}}_\mathrm{eq} = \tr(\hat{\rho}_\mathrm{eq} \hat{A})$ of $\hat{A}$ is therefore given by the expression
\begin{equation}\label{eq:expValPIMD}
    \braket{\hat{A}}_\mathrm{eq} = \frac{1}{Z} \lim_{P \to +\infty} \, \int_\R \, \prod_{i=1}^{N} \diff^3 \bfp_i^{(1)} \dotsm \diff^3 \bfp_i^{(P)} \int_{D(L)} \, \prod_{i=1}^{N} \diff^3 \bfr_i^{(1)} \dotsm \diff^3 \bfr_i^{(P)} \, \Bigl[A_P \bigl( \bigl\{\bfr_i^{(j)}\bigr\} \bigr) \, \ee^{- \beta H_P}\Bigr] \ .
\end{equation}
Using the approximate partition function $Z_P$ introduced above, we can define an approximation for $\braket{\hat{A}}_\mathrm{eq}$ of accuracy $P$ by setting
\begin{equation*}
    \braket{\hat{A}}_P \ce \frac{1}{Z_P} \int_\R \, \prod_{i=1}^{N} \diff^3 \bfp_i^{(1)} \dotsm \diff^3 \bfp_i^{(P)} \int_{D(L)} \, \prod_{i=1}^{N} \diff^3 \bfr_i^{(1)} \dotsm \diff^3 \bfr_i^{(P)} \, \Bigl[A_P \bigl( \bigl\{\bfr_i^{(j)}\bigr\} \bigr) \, \ee^{- \beta H_P}\Bigr] \ ,
\end{equation*}
and this approximation satisfies $\braket{\hat{A}}_\mathrm{eq} = \lim_{P \to + \infty} \braket{\hat{A}}_P$. In terms of an MD trajectory, the quantity $\braket{\hat{A}}_P$ can be estimated by sampling the function $A_P$ from \cref{eq:classicalEstimator} over a trajectory generated by the classical polymer-ring Hamiltonian $H_P$:
\begin{equation}\label{eq:MDapproximation}
    \braket{\hat{A}}_P \approx \frac{1}{M} \sum_{k=1}^{M} \left[ \frac{1}{P} \sum_{j=1}^{P} A \bigl(\bfr_1^{(j)} (k \Delta t), \dotsc, \bfr_N^{(j)} (k \Delta t)\bigr) \right] \ ,
\end{equation}
where $\Delta t$ is the time step of the simulation and $M \Delta t$ the total temporal length of the trajectory. (That is, the observable $A$ is calculated at each point $\bfr_i^{(j)} (k \Delta t)$ of the discretized trajectory.) Choosing $P$ large enough, \cref{eq:MDapproximation} gives an approximation for $\braket{\hat{A}}_\mathrm{eq}$ as well.

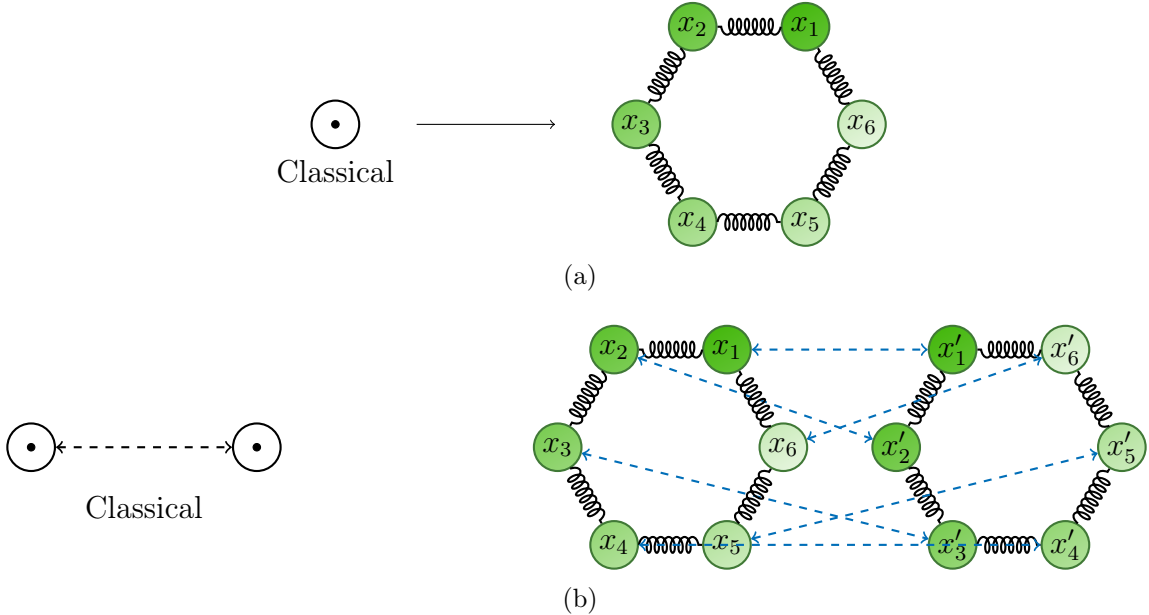
\begin{figure}
    \centering
    \subfloat[]{%
        \begin{tikzpicture}

            \foreach \a/\t/\b [count=\n] in {60/70/60, 120/60/50, 180/50/40, 240/40/30, 300/30/20, 360/20/10}
                \node [circle, thick, minimum size=18pt, inner sep=-2, draw=nodeGreen!90!black, top color=nodeGreen!\t, bottom color=nodeGreen!\b, shading angle=30] (q\n) at (\a:\reflen) {$x_{\n}$};
            
            \foreach \n [count=\i] in {2, 3, 4, 5, 6, 1}
                \draw [spring] (q\i) -- (q\n);
        
            \node [left of=q3, xshift=-3cm, circle, draw, fill, inner sep=1pt] {};
            \node [left of=q3, xshift=-3cm, circle, draw, thick, minimum size=18pt, inner sep=-2, label={below: Classical}] (cl1) {};
            \draw [->] ([xshift=0.75cm]cl1.east) -- ([xshift=-0.75cm]q3.west);
        
        \end{tikzpicture}
    } \\
    \subfloat[]{%
        \begin{tikzpicture}

            \foreach \a/\b/\u/\l [count=\n] in {60/120/70/60, 120/180/60/50, 180/240/50/40, 240/300/40/30, 300/360/30/20, 360/60/20/10}
                {
                \node [circle, thick, minimum size=18pt, inner sep=-2, draw=nodeGreen!90!black, top color=nodeGreen!\u, bottom color=nodeGreen!\l, shading angle=30] (q\n) at (\a:\reflen) {$x_{\n}$};
                \node [xshift=4.5cm, circle, thick, minimum size=18pt, inner sep=-2, draw=nodeGreen!90!black, top color=nodeGreen!\u, bottom color=nodeGreen!\l, shading angle=30] (p\n) at (\b:\reflen) {$x_{\n}^\prime$};
                }
            
            \foreach \n [count=\i] in {2, 3, 4, 5, 6, 1}
                {
                \draw [spring] (q\i) -- (q\n);
                \draw [spring] (p\i) -- (p\n);
                \draw [<->, dashed, thick, color=intBlue] (q\i) -- (p\i);
                }
        
            \foreach \x [count=\i] in {-3cm, -6cm}
                {
                \node [left of=q3, xshift=\x, circle, draw, fill, inner sep=1pt] {};
                \node [left of=q3, xshift=\x, circle, draw, thick, minimum size=18pt, inner sep=-2] (cl\i) {};
                }
        
            \draw [<->, dashed, thick] (cl2) -- (cl1) node [midway, below, yshift=-0.5cm] {Classical};
        \end{tikzpicture}
    }
    \caption{Panel (a) illustrates the mapping from the classical representation of an atom as a spherical object to the quantum representation as a polymer ring whose beads are connected by harmonic oscillators, here for $P = 6$. Panel (b) shows how interactions are modeled in the polymer ring representation, here for $N = 2$ particles and $P = 6$: only beads with the same index on different polymer rings interact with each other.}
    \label{fig:pimd}
\end{figure}

With the above method, the problem of computing equilibrium expectation values of quantum-mechanical observables is mapped to an effective sampling via molecular dynamics in the space of interacting polymer rings (see \cref{fig:pimd}). While the dynamics is fictitious, the spatial sampling of the conformations is statistically equivalent to the quantum statistics of interacting particles; in particular, quantum effects arise because, contrary to standard classical MD, the particles we have considered are not rigid spheres but space-delocalized entities due to the fluctuating shape of each ring. It should be noted that instead of employing polymer rings and their dynamics, one may equivalently use the formalism of centroid molecular dynamics \cite{Jang1999} which uses quasi-classical objects, each defined as an average over all the beads of one polymer ring as described before.

While the PIMD method allows us to calculate, via classical MD, quantum ensemble averages of physical observables in equilibrium, it does not tell us, however, how to treat these averages in situations out of equilibrium. For this one requires a non-equilibrium MD technique and its quantum equivalent. In the next section, we will discuss such a classical MD method known in the literature, and we will argue how it can be consistently translated to the quantum formalism. It must be reported that, although the time evolution is fictitious, there are variations of PIMD that allow for the calculation of time-dependent properties in the form of a linear response \cite{Jang1999, Habershon2013}; we will come back to the connection of our work with such methodologies later on.

\section{Going beyond equilibrium}\label{sec:beyondEquilibrium}

A specific feature of non-equilibrium statistical mechanics is that the probability density in phase space for classical systems, or the density matrix for quantum system, is not known \textit{a priori} and cannot be related to any of the expressions for the equilibrium ensembles (microcanonical, canonical, grand canonical, and their variations) \cite{Tuckerman2023, Huang1987}. Consequently, the problem is that one does not know how to calculate statistical averages of physical observables in out-of-equilibrium situations. Since the equilibrium distribution is usually known, one can often treat the non-equilibrium situation as a first-order perturbation or response \cite{Blair2024}. In MD specifically, many refined methods are known to overcome this problem and treat systems out of equilibrium \cite{Ciccotti2005}; here we will describe a particular one, called dynamical non-equilibrium molecular dynamics (D-NEMD), which was originally developed by Ciccotti and Jacucci \cite{Ciccotti1975}, and successfully applied to various out-of-equilibrium situations \cite{Bonella2017, EbrahimiViand2020, Wang2014, Ferrario2025}. This method requires only the equilibrium distribution (or even a stationary distribution) to determine non-equilibrium statistical averages of physical observable. Moreover, it even allows going beyond linear response theory (see, e.g., Ref. \cite{Wang2014}).

Importantly for us, it turns out that the classical D-NEMD formalism can be extended to quantum systems by utilizing the equivalence of the Schrödinger and the Heisenberg representation. This extension, together with the relation between classical MD and the calculation of quantum statistical averages through the PIMD method discussed in \cref{sec:PIMD}, allows connecting the time-dependent density matrix $\rho(t)$, which is a solution of the Lindblad equation \eqref{eq:Lindblad}, with the path integral treatment of particles, leading to an indirect solution of the former. This connection is the main novelty proposed in this work.

\subsection{Classical D-NEMD method}

In classical physics, the time-dependent statistical average of an observable $A = A (\bfr, \bfp)$ in a state $\varrho(t) = \varrho (\bfr, \bfp, t)$, which is the solution of the Liouville equation $\dot{\varrho} = - \{H, \varrho\}$ for some time-dependent Hamiltonian $H = H (\bfr, \bfp, t)$, is given by
\begin{equation}\label{eq:classicalDNEMD}
    \braket{A}(t) = \int_\Gamma A(\bfr, \bfp) \, \varrho(\bfr, \bfp, t) \diff \mu = \braket[\big]{A, \varrho(t)}_{L^2} \ ,
\end{equation}
where $\Gamma = \R^{6N}$ denotes the phase space spanned by positions $\bfr = (x_1, \dotsc, x_{3N})$ and momenta $\bfp = (p_1, \dotsc, p_{3N})$, $\diff \mu = \prod_{i=1}^{3N} \diff q_i \diff p_i$ is the Liouville measure on $\Gamma$, and $\braket{\cdot,\cdot}_{L^2}$ denotes the usual inner product on the space $L^2(\Gamma, \mu)$. The starting point for the D-NEMD method of Ciccotti and Jacucci \cite{Ciccotti1975} is the observation that $\braket{A}(t)$ can also be written as an average of the time-dependent observable $A(t) \ce A \bigl(\bfr(t), \bfp(t)\bigr)$, where $t \mapsto \bfr(t)$ and $t \mapsto \bfp(t)$ solve Hamilton's equations of motion, with respect to the equilibrium distribution $\varrho_\mathrm{eq} \ce \varrho (\bfr, \bfp, 0)$. Indeed, noting that the non-equilibrium density can be expressed as $\varrho (\bfr, \bfp, t) = U^\dagger (t, 0) \, \varrho_\mathrm{eq}$, where $U(t, 0) \ce \exp \bigl(- \ii \int_0^t L(t^\prime) \diff t^\prime\bigr)$ and $\ii L \ce - \{H, \,\cdot\,\}$ are the classical time evolution operator and Liouvillian, respectively, and that the time-dependent observable is given by $A \bigl(\bfr(t), \bfp(t)\bigr) = U (t, 0) \, A (\bfr, \bfp)$ \cite{Wang2014, Ciccotti1979}, it follows that
\begin{equation}\label{eq:DNEMD}
    \braket{A}(t) = \braket[\big]{A, U^\dagger (t, 0) \, \varrho_\mathrm{eq}}_{L^2} = \braket[\big]{U (t, 0) \, A, \varrho_\mathrm{eq}}_{L^2} = \braket[\big]{A(t), \varrho_\mathrm{eq}}_{L^2} \ec \braket[\big]{A(t)}_\mathrm{eq} \ .
\end{equation}

The concrete D-NEMD protocol that uses \cref{eq:DNEMD} to calculate $\braket{A} (t)$ can be summarized as follows (see \cref{fig:d-nemd} for an illustration): first, one has to generate long equilibrium trajectories in order to sample the distribution $\varrho_\mathrm{eq}$, represented by the solid line at the top of \cref{fig:d-nemd}. One then takes uncorrelated points, labeled $t_0, \dotsc, t_n$ in \cref{fig:d-nemd}, along such trajectories and initiates, for each of these points separately, the MD algorithm with the full non-equilibrium Hamiltonian $H$. This leads to several non-equilibrium branch trajectories, represented by the solid vertical lines in \cref{fig:d-nemd}, each starting at a point of equilibrium. The expectation value of the observable $A$ after $m$ time steps is then calculated by averaging over all the values that $A$ takes at the points $t_m^{b_k}$, $0 \le k \le n$, in each of the non-equilibrium branch trajectories:
\begin{equation*}
    \braket{A} (t_m) = \frac{1}{n+1} \sum_{k=0}^{n} A \bigl(t_m^{b_k}\bigr) \ .
\end{equation*}

\begin{figure}[t]
    \scalebox{0.8}{%
    \begin{tikzpicture}[font=\Large]
        \foreach \x/\y [count=\i from 0] in {0/0, 3/1, 6/0.25, 9/0.5, 12/-0.25, 15/0.25, 18/0}
        {
            \coordinate (t\i) at (\x,\y);
        }
        \foreach \i in {0, 1, 2, 3, 6}
        {
            \ifthenelse{\NOT \i = 6}
            {
                \node [circle, fill, inner sep=2pt] at (t\i) {};
                \node [above, yshift=0.1cm] at (t\i) {$t_{\i}$};
            }
            {
                \node [circle, fill, inner sep=2pt] at (t\i) {};
                \node [above, yshift=0.1cm] at (t\i) {$t_n$};
            }
        }
        \draw [very thick] (t0) .. controls (1,1) and (2,1.25) .. (t1)
                                .. controls (4,0.75) and (5,0.25) .. (t2)
                                .. controls (7,0.25) and (8,0.5) .. (t3)
                                .. controls (10,0.5) and (11,-0.25) .. (t4)
                                .. controls (13,-0.25) and (14,0.25) .. (t5)
                                .. controls (16,0.25) and (17,0) .. (t6);
        \foreach \i/\x in {0/0, 1/3, 2/6, 3/9, n/18}
        {
            \foreach \j/\y in {1/-2, 2/-4, 3/-6, 4/-8}
            {
                \coordinate (t\j b\i) at (\x,\y);
                \ifthenelse{\NOT \j = 4}
                {
                    \node [circle, fill, inner sep=2pt] at (t\j b\i) {};
                    \node [above right] at (t\j b\i) {$t_{\j}^{b_{\i}}$};
                }
                {}
            }
        }
        \foreach \i/\k in {0/0, 1/1, 2/2, 3/3, n/6}
        {
            \draw [trajectory] (t\k) -- (t1b\i);
            \foreach \j [count=\n] in {2, 3, 4}
            {
                \draw [trajectory] (t\n b\i) -- (t\j b\i);
            }
        }
        \draw [dashed] (t1b0) -- (t1bn);
        \draw [dashed] (t2b0) -- (t2bn);
        \draw [dashed] (t3b0) -- (t3bn);
        \foreach \y in {0.5, -1.5, -3.5, -5.5}
        {
            \node at (13.5,\y) {$\hdots$};
        }
        \foreach \x in {0.5, 3.5, 6.5, 9.5, 18.5}
        {
            \node at (\x,-7) {$\vdots$};
        }
        \node [font=\large] at (9,-9) {$\big\vert t_i - t_{i+1} \big\vert = \Delta t$ and $\big\vert t_i - t_1^{b_i} \big\vert, \big\vert t_i^{b_j} - t_{i+1}^{b_j} \big\vert = \Delta t^\prime$ for all $0 \le i, j \le n$.};
    
    \end{tikzpicture}}
    \caption{Illustration of the D-NEMD protocol. The horizontal curve at the top represents the equilibrium trajectory. The vertical curves, originating from temporally equidistant points along the former, represent trajectories where the external perturbation acts on the particles. The non-equilibrium statistical average of a physical observable is calculated using the values of the observable on points of the branched trajectory on the horizontal dotted lines.}
    \label{fig:d-nemd}
\end{figure}

\subsection{Extension to quantum systems}

In the quantum-mechanical equivalent of \cref{eq:classicalDNEMD}, the expectation value of the operator $\hat{A}$ is evaluated in the Schrödinger picture, where the density operator $\hat{\rho}$ is time-dependent and the observable is time-independent, see also \cref{eq:timeDependentExpectation}:
\begin{equation*}
    \braket{\hat{A}} (t) = \tr \bigl(\hat{\rho}(t) \hat{A}\bigr) \ .
\end{equation*}
To obtain an expression analogous to \eqref{eq:DNEMD}, one may use the well-known fact (see, e.g., \cite[p. 132]{Pillet2006}) that the Schrödinger representation can be transformed into the Heisenberg representation with the help of the time evolution operator $\hat{U}(t, 0) = \MT \exp \bigl(- \frac{\ii}{\hbar} \int_0^t \hat{H}(t^\prime) \diff t^\prime\bigr)$: it holds that $\hat{\rho}(t) = \hat{U}(t, 0) \, \hat{\rho}(0) \, \hat{U}^\dagger (t, 0)$ and $\hat{A}(t) = \hat{U}^\dagger (t, 0) \, \hat{A} \, \hat{U} (t, 0)$, hence cyclicity of the trace gives \footnote{From a structural point of view, the quantum-mechanical argument is analogous to the classical argument of \cref{eq:classicalDNEMD} because the trace gives rise to an inner product on the space of Hilbert-Schmidt operators; see, e.g., Ref. \cite[Section II, esp. Eq. (II.13)]{Bach2000}.}
\begin{equation}\label{eq:expValTransform}
    \braket{\hat{A}} (t) = \tr \bigl(\hat{U}(t, 0) \, \hat{\rho}(0) \, \hat{U}^\dagger (t, 0) \, \hat{A}\bigr) = \tr \bigl(\hat{\rho}(0) \, \hat{U}^\dagger (t, 0) \, \hat{A} \, \hat{U}(t, 0)\bigr) = \tr \bigl(\hat{\rho}(0) \hat{A}(t)\bigr) \ .
\end{equation}
The term on the right-hand side is the quantum equivalent of the classical expression of Eq. \eqref{eq:DNEMD} which can be calculated through the D-NEMD procedure, as discussed before. It follows that if we can map the right-hand side of Eq. \eqref{eq:expValTransform} onto classical quantities, then we would actually obtain $\braket{\hat{A}} (t)$ using classical tools, namely, molecular dynamics. To this end, notice that in the PIMD formalism (i.e., with an effective classical treatment of a quantum system), the equilibrium density operator $\hat{\rho}(0)$ is replaced by its position-space matrix elements $\rho(x, x) = \braket{x \,\vert\, \hat{\rho}(0) \,\vert\, x}$, which are calculated by sampling polymer-ring configurations using classical MD (cf. \cref{sec:PIMD}). Thus, combining the D-NEMD with the PIMD method, one could efficiently calculate time-dependent statistical averages of observables by sampling the equilibrium distribution with PIMD and the time-dependent observable via the D-NEMD approach.

To give consistency to the approach proposed above, first and foremost we need to write the time-dependent observable $\hat{A}(t)$, acting as an operator in the underlying Hilbert space, as an effective classical object that can be calculated through D-NEMD, that is, as a real-valued function $A \bigl(\bfr_1^{(j)}(t), \dotsc, \bfr_N^{(j)}(t)\bigr)$, $j \in \set{1, \dotsc, P}$, as in \cref{eq:classicalEstimator} in the PIMD formalism. Formula \eqref{eq:classicalEstimator} is certainly valid under the assumption of stationary equilibrium, hence it is necessary at this point to argue that $A \bigl(\bfr_1^{(j)}(t), \dotsc, \bfr_N^{(j)}(t)\bigr)$ is well-defined also along trajectories out of equilibrium. This means that we need to justify that $A \bigl(\bfr_1^{(j)}(t), \dotsc, \bfr_N^{(j)}(t)\bigr)$ is dynamically equivalent, in classical effective terms, to the time-dependent observable $\hat{A}(t)$, acting as an operator in Hilbert space. The argument that we propose is the following: the exact time evolution of $\hat{A}(t)$ is governed by Heisenberg's equation of motion,
\begin{equation}\label{eq:Heisenberg}
    \frac{\diff \hat{A}(t)}{\diff t} = \frac{\ii}{\hbar} \, \bigl[ \hat{H}(t), \hat{A}(t) \bigr] \ .
\end{equation}
The application of the path integral formalism implies that $A \bigl(\bfr_1^{(j)}(t), \dotsc, \bfr_N^{(j)}(t)\bigr)$ is the projections of $\hat{A}(t)$ onto position space at any time $t$:
\begin{equation}\label{eq:positionSpaceProjection}
    A \bigl(\bfr_1^{(j)}(t), \dotsc, \bfr_N^{(j)}(t)\bigr) = \dbraket[\big]{\bfr_1^{(j)}, \dotsc, \bfr_N^{(j)}}{\hat{A}(t)}{\bfr_1^{(j)}, \dotsc, \bfr_N^{(j)}} \ .
\end{equation}
Since $\hat{A}(t)$ satisfies \cref{eq:Heisenberg}, one can conclude that the expression \eqref{eq:positionSpaceProjection} entering the path integral can be used in the D-NEMD scheme because it represents the effective dynamical analog of $\hat{A}(t)$ in classical terms, thus it is dynamically defined also along trajectories out of equilibrium; in PIMD, the polymer-ring trajectories out of equilibrium represent the departure of ($\bfr_1^{(j)}(t), \dotsc, \bfr_N^{(j)}(t))$ from the equilibrium trajectory (generated by the Hamiltonian $H_{P}$ from \cref{eq:effectiveQuantumH}) due to the action of the dissipative term in the Hamiltonian. Thus, the application of the D-NEMD PIMD procedure to the problem of calculating $\tr \bigl(\hat{\rho}(0) \hat{A}(t)\bigr)$ is physically justified. We will refer to this combined D-NEMD PIMD method for computing expectation values $\braket{\hat{A}} (t)$ out of equilibrium as \emph{NPI method} for brevity (non-equilibrium path integral). In conclusion, any ensemble average $\braket{A}(t) = \tr \bigl(\rho(t) A)$ obtained via the time-dependent density operator $\rho (t)$ from the Lindblad equation is consistently close (for large but finite $P$), or even equivalent (for $P \to + \infty$), to the expectation value $\tr \bigl(\rho(0) A(t)\bigr)$ obtained via the NPI method.

The limitations of the PIMD method compared to the Lindblad equation must always be remembered. While the latter captures the exact reduced dynamics of an open system (given all the necessary approximations needed to derive the equation), the former can only calculate ensemble averages which inevitably neglects short-time quantum coherence; indeed, since in this framework one computes the statistical average of the time-dependent observable, one effectively performs different independent measurements of the observable and averages over them. However, as underlined in Ref. \cite{Habershon2013}, rapid quantum decoherence is a characteristic of many process in condensed matter systems, so the technique is justified for a very large class of systems. A proof of the conceptual consistency of the theoretical framework that we have proposed here is the following observation: in the limit of equilibrium, the computation of time correlation functions within the NPI method automatically reduces to the same procedure routinely used in PIMD studies of equilibrium \cite[Eq. (11)]{Habershon2013}; we provide the explicit argument in Appendix \ref{app:limitEquilibrium}.

\subsection{\texorpdfstring{Positivity of $\rho(t)$}{Positivity of rho(t)}}

From the discussion above, one might get the impression that NPI simulations are independent of the Lindblad equation, in that they provide an alternative computational method to the latter for large systems and long time scales, where the Lindblad equation is of no practical interest. However, as indicated before, this is not the case since the Lindblad equation is crucial for assuring positivity of the density operator $\hat{\rho}(t)$ at all times $t$. More precisely, if one deals with an open quantum system whose coupling to the external environment (e.g., thermostats) can be written in the Lindblad form \eqref{eq:Lindblad} (or its classical analog), then the time-dependent expectation values $\braket{\hat{A}} (t)$ calculated using the equivalence
\begin{equation*}
    \tr \bigl(\hat{\rho}(t) \hat{A}\bigr) = \tr \bigl(\hat{\rho}(0) \hat{A}(t)\bigr) \ ,
\end{equation*}
and the NPI method are physically consistent because, as mentioned above, the Lindblad equation assures positivity of the density operator $\hat{\rho}(t)$ throughout the entire time evolution. Conversely, if the coupling to the external reservoir is not of the form \eqref{eq:Lindblad}, one cannot be assured that computing $\tr \bigl(\hat{\rho}(0) \hat{A}(t)\bigr)$ via NPI is well-posed since $\hat{\rho}(t)$ is not necessarily positive for every $t$ and hence might have negative eigenvalues in which case the expression $\tr \bigl(\hat{\rho}(t) \hat{A}\bigr)$ cannot be interpreted as a physically meaningful expectation value since $\hat{\rho}(t)$ does not give rise to a quantum probability measure as it will generally lead to negative probabilities.

In essence, therefore, the interesting formal result is that the possibility of mapping the external source in the NPI method onto a Lindbladian is a necessary condition for using this method to calculate statistical averages of physical observables in situations out of equilibrium. In PIMD calculations previously presented in the literature, the positivity of the density operator $\hat{\rho}(t)$ has not been a concern because, until now, only systems in equilibrium or at worst near equilibrium have been treated. Thus, the novelty put forward in this work is the use of NPI for situations out of equilibrium, with a necessary condition to settle the question about the positivity of $\hat{\rho}(t)$, as this condition enforces the internal physical consistency of NPI simulations.

To make this argument more precise, suppose that the coupling of the open system $S$ to the external environment $R$ entering the NPI method satisfies the Born-Markov approximation and is thus described by the general Redfield master equation
\begin{equation}\label{eq:Redfield}
    \frac{\diff \hat{\rho}_S(t)}{\diff t} = - \left(\frac{\alpha}{\hbar}\right)^2 \, \int_{0}^{+\infty} \tr_R \Bigl[ \bigl[ \hat{H}_\mathrm{int}(t), [\hat{H}_\mathrm{int}(t - t^\prime), \hat{\rho}_S(t) \otimes \hat{\rho}_R] \bigr] \Bigr] \diff t^\prime \ .
\end{equation}
Here, the parameter $\alpha \in \R$ determines the strength of the interaction (see Appendix \ref{app:Lindblad} for details). Suppose, furthermore, that the secular approximation cannot be applied to this system, meaning that the Redfield equation cannot be written in Lindblad form (see again Appendix \ref{app:Lindblad} for further explanations). It is well-known that in this case, there can exist times $t$ for which the solution $\rho_S(t)$ of \cref{eq:Redfield} is not a positive operator \cite{Davies1974, Davies1976, Dümcke1979, Suarez1992, Benatti2005, Whitney2008, Tupkary2023, Campaioli2024}, hence does not describe a proper physical state. Nevertheless, the Redfield equation can be written in the form \cite[Eqs. (96) and (102)]{Benatti2005}, \cite[Eq. (8)]{Tupkary2023}
\begin{equation*}
    \frac{\diff \hat{\rho}_S(t)}{\diff t} = - \frac{\ii}{\hbar} \, [\hat{H}, \hat{\rho}_S] + \wt{\MD} \bigl(\hat{\rho}_S(t)\bigr) \ ,
\end{equation*}
which shows formal similarity to the Lindblad equation, cf. \cref{eq:Lindblad}, where the dissipative term $\wt{\MD} \bigl(\rho_S(t)\bigr)$, however, is not as simple as in \cref{eq:LindbladDissipative}. Thus, the application of the NPI procedure, with an external source as described above, would proceed in the same way as if the external coupling were of Lindblad form, and would deliver numerical results. However, since the Redfield equation does not guarantee positivity of $\hat{\rho}_S (t)$, the NPI method may deliver physically inconsistent results. Such a possible inconsistency could be discovered only by explicitly solving the Redfield equation for the system under investigation to check whether the density operator is positive or not; hence, the method does not yield a computational simplification as the complicated master equation has to be solved anyway. In contradistinction to this, if the system is known to satisfy the secular approximation, \cref{eq:Redfield} can be simplified to the Lindblad equation \eqref{eq:Lindblad} which assures positivity of the density operator, and thus, in this case, one can be certain that NPI simulations are physically consistent. In other words, it is sufficient to know that the coupling of the system to the reservoir is described by a term of the form \eqref{eq:LindbladDissipative} in order to conclude that the NPI method yields reliable results. Summarizing, we have justified the following

\begin{conclusion}
    Consider an open quantum system coupled to an external reservoir such that the secular approximation (or rotating wave approximation) is valid. Then the time-dependent, non-equilibrium expectation values calculated via NPI are physically consistent.
\end{conclusion}

It is very interesting to note that, paradoxically, even though one can bypass the task of obtaining an explicit solution to the Lindblad equation, the latter plays a key role \textit{per se} for establishing physical consistency of the numerical NPI method, while the Redfield equation, for example, does not offer the possibility to draw such conclusions and instead requires an explicit case-by-case verification of the density operator.

\begin{remark}
    It is important to note that in recent years, the violation of positivity in the Redfield equation as well as the necessity of the secular approximation for the transition from the Redfield to the Lindblad form has been investigated thoroughly; see, for example, Refs. \cite{Gaspard1999, Schaller2008, Taj2008, Whitney2008, Benatti2010, Majenz2013, Jeske2015, Farina2019, Hartmann2020, Trushechkin2021, Tupkary2022}. This has led to certain alternatives to the secular approximation, which also give rise to a master equation that preserves positivity of its solutions. Therefore, in more generality (but also less concrete terms), we can state that a sufficient condition for the physical consistency of the NPI method is that the coupling to the external reservoir yields, through an appropriate approximation procedure, a completely positive time-evolution of the reduced density operator.
\end{remark}

Explicit examples of external sources that are classical analogs of those for which the coupling term can be expressed in Lindblad form are Drude and Ohmic baths, written via harmonic oscillators, which can be rewritten in the form of a Langevin bath (see, e.g., Refs. \cite{Diosi1993,Das2020,Chen2025} and references therein). As suggested above, sources of this type can be used to create typical situations out of equilibrium which are of physical and technological interest, namely generating a heat flux through a thermal gradient by coupling different regions of the system to different baths.

One should note that the concepts developed here can also be used in situations where the number of particles in the systems is not fixed (see, e.g., Ref. \cite{Reible2025pre}); in fact, the PIMD method has already been tested for a system with variable particle number, as explicitly discussed in Ref. \cite{DelleSite2024jpa}.

\subsection{Accuracy}\label{subsec:accuracy}

Regarding the results above, it is clear that they only approximate the true quantum-mechanical non-equilibrium expectation value and that the quality of the approximation is controlled by the number $P$ of beads in the polymer ring; in fact, as noted in \cref{sec:PIMD},
\begin{equation*}
  \lim_{P \to +\infty} \braket{\hat{A}}_P (t) = \braket{\hat{A}} (t) \ .
\end{equation*}
That is, only in the limit of infinitely many beads is the full quantum-mechanical equivalence justified. This means that in the NPI method, for a sufficiently large value of $P$ we have $\braket{\hat{A}} (t) \approx \braket{\hat{A}}_P (t)$. From the practical point of view, the NPI method suggests that
\begin{equation}\label{mdquantnoneq}
    \braket{\hat{A}}_P (t_m) = \frac{1}{n + 1} \sum_{k=0}^{n} \frac{1}{P} \sum_{j=1}^{P} A \bigl(\bfr_1^{(j)} (t_m^{b_k}), \dotsc, \bfr_N^{(j)} (t_m^{b_k})\bigr) \ ,
\end{equation}
where $n$ is the number of branched trajectories, thus each term of this sum is calculated at a time $t_m^{b_k}$ of each of the branched trajectories (cf. \cref{fig:d-nemd}). Moreover, for physical consistency, one should carry out a study increasing $P$ systematically until results numerically converge so that one can be sure to be in an acceptable range of validity of the relation $\lim_{P \to +\infty} \braket{\hat{A}}_P (t) = \braket{\hat{A}} (t)$, and thus to capture the quantum properties of the system. By calculating the quantity in \cref{mdquantnoneq} at different values of $t_m$ along the equilibrium trajectory, one builds a time series that reflects the time evolution of the observable $\hat{A}$ out of equilibrium.

\section{Application of the method}\label{sec:application}

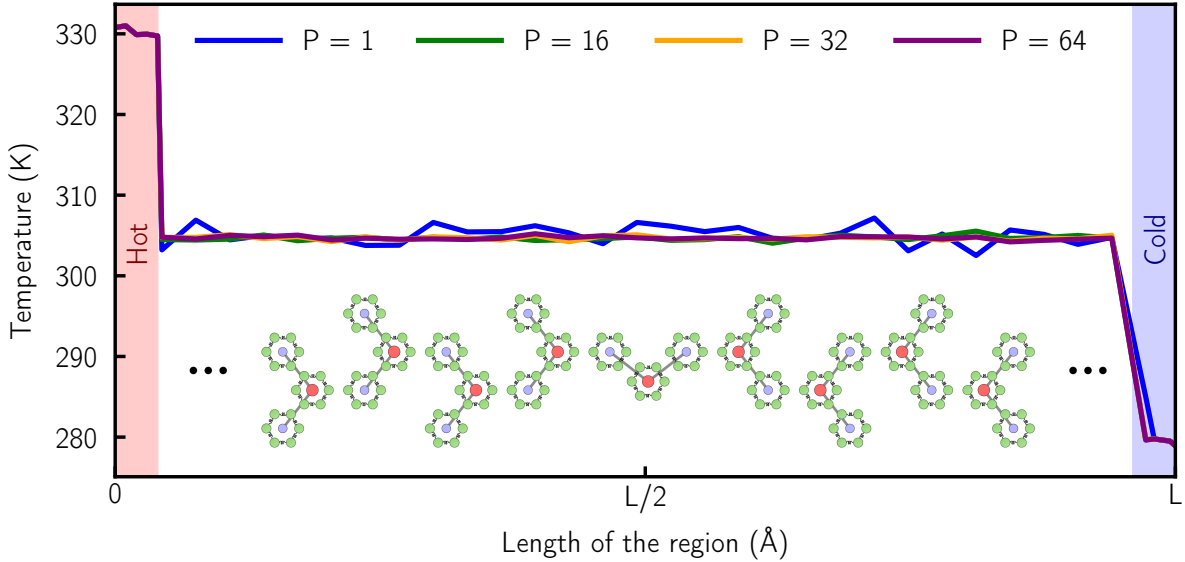
\begin{figure}
    \centering
    \scalebox{0.65}{\input{figures/thermal_profiles.pgf}}
    \caption{Temperature profile of a one-dimensional chain of 87 water molecules, each one modeled as a polymer ring (pictorially visualized above the $x$-axis), in a thermal gradient for different values of the number of beads $P$. On the left-hand side of the chain is a \enquote{hot} temperature reservoir at $330 \, \text{K}$, and on its right-hand side is a \enquote{cold} reservoir at $280 \, \text{K}$. As expected, the chain reaches a steady state with a uniform temperature at around $305 \, \text{K}$, and the higher the value of $P$, the smaller the deviation of the temperature along the chain from this steady-state value.}
    \label{fig:thermalProfile}
\end{figure}

As motivated in the introduction, we have chosen a one-dimensional chain of water molecules in a thermal gradient as an illustrative application of the proposed methodology. Technical details of the simulation can be found in Appendix \ref{app:techdet}, while a pictorial illustration of the system can be found above the $x$-axis in \cref{fig:thermalProfile}.

As previously discussed, a system as simple as the one we are treating here would not be accessible by the Lindblad approach, unless one has very large computational facilities available, or one maps the problem onto a highly coarse-grained model where the chemical details are neglected, e.g., a chain of harmonic oscillators. Instead, the essential quantum nature of the molecular atoms (especially the hydrogen), namely spatial delocalization, can be represented well by the PI theory and its numerical implementation in terms of interacting polymer rings. The question that arises concerns the applicability of the proposed method, that  is, whether there is convergence of the calculated physical quantities of interest for finite $P$. If the answer is positive, then one may ask which valuable physical knowledge one gains that is not accessible through the Lindblad approach, because of the computational resources or excess of model simplification, and/or by standard classical models because they are not optimally designed for such properties.

\begin{figure}
    \centering
    \scalebox{0.65}{\input{figures/flux.pgf}}
    \caption{Heat flux along the one-dimensional chain of water molecules for different values of the bead number $P$. The data illustrates that for $P = 64$, the heat flux has essentially converged.}
    \label{fig:heatFlux}
\end{figure}
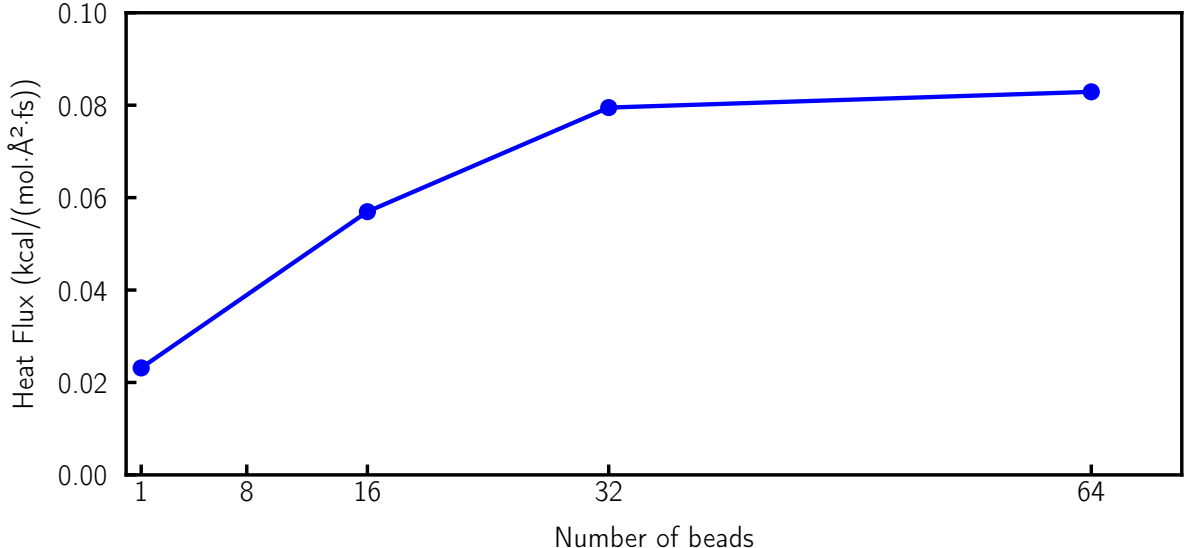

A one-dimensional chain of uniform elements in a thermal gradient will reach a steady state in which the temperature is uniform along the chain at the average value between the two temperatures of the hot and cold reservoir \cite{carstenmolphys}. Thus, a first test of validity of the method consists in simulating the system at different values of $P$, from $P=1$ (classical limit) to systematically larger values, in order to show that all the simulations lead to the expected steady state. Such a test will show consistency of the method because it gives the physically expected result for a reference situation out of equilibrium, above all for large values of $P$; as reported in \cref{fig:thermalProfile}, such a study indeed shows the expected behavior. Next, we need to address the question whether a physical observable of interest in this context, calculated in the non-equilibrium steady state, converges for a certain (finite and numerically acceptable) value of $P$. If the convergence could not be found, then this would imply that out of equilibrium, the method would only work in the ideal situation of $P$ being very large, and thus the method would not be numerically convenient. We have chosen the average heat flux along the chain as a physical observable, and \cref{fig:heatFlux} shows that for $P=64$ this quantity has essentially converged; note that for liquid water in equilibrium, usually $P=32$ is sufficient (see, e.g., Ref. \cite{Agarwal2015}).

The final question concerns the overall utility of the approach, that is, whether one can detect a quantum signature in the physical observable; this would be the added value of the proposed approach compared to the standard full quantum and full classical techniques. In a chain of water molecules, one expects the heat flux to increase when the quantum character of the atoms moves from the classical case ($P=1$) to higher and higher values of $P$. In fact, due to the delocalization of atoms in space, one expects a higher probability for the atom of one molecule to bond with atoms of neighboring molecules (e.g., O--H bond in water), thereby transporting the heat towards the colder atoms. Such an outcome would also be consistent with the results obtained in near-equilibrium studies of water, where an enhanced flexibility of the O--H bond, compared to the classical case, was detected by the PIMD method \cite{frontchem}. Figure \ref{fig:heatFlux} shows that this is indeed the case, and thus we gain the interesting information about the enhanced heat flux in the chain of water molecules, due to the quantum character of the atoms as described by the PI model.

We can conclude that our basic test gives satisfactory and encouraging results, and thus stimulates us to apply the method to more complex systems tracing its range of applicability and finding paths on technical and conceptual improvement.

\section{Conclusions}

The comparison of the Lindblad approach with the PIMD approach to open quantum systems has suggested the extension of the latter to situations out of equilibrium. We have formulated a physically sound and, as much as possible within necessary assumptions,  consistent theoretical framework for such an extension. In addition to the general formulation, we have found a necessary condition that assures the positivity of the density matrix also in the PIMD formulation by leveraging structural properties of the Lindblad equation. In PIMD such a condition applies to the choice of the system--environment coupling (i.e., to external sources), which must be formally consistent with the expression of the Lindbladian. A typical example of consistency is the use of Langevin thermostats, such as, for example, in the creation of a gradient of temperature and thus in the study of heat transport in chemo-physical systems. The extension of PIMD to non-equilibrium situations paves the way for studying a large class of systems at atomic and molecular scale. The proposed method goes beyond current PIMD-like approaches in the sense that the latter can describe situations out of equilibrium, such as adiabatic reactions, only in terms of linear response through the Kubo transform correlation functions \cite{instantons1,instantons2,Habershon2013}, while our approach can treat cases beyond the linear response, see e.g., Ref. \cite{Wang2014}. Finally, the possibility of performing such simulations can yield data to describe coarser models for the direct treatment in the Lindblad equation and, more generally, future ideas based on specific chemical compositions for quantum materials and the corresponding technology.

\acknowledgments{This work was supported by the DFG Collaborative Research Center 1114 ``Scaling Cascades in Complex Systems'', Project No. 235221301, Project C01 ``Adaptive coupling of scales in molecular dynamics and beyond to fluid dynamics'' (L.D.S. and B.M.R). Further support by the DFG, Project No. DE 1140/15-1, ``Mathematical Modeling and computational implementation of open quantum systems for molecular simulations'', is acknowledged (L.D.S. and S.A.).}

\appendix
\section{Derivation of the Redfield and Lindblad equation}\label{app:Lindblad}

In the following, we provide a short outline of the derivation of the Redfield and Lindblad equation; see Refs. \cite{BreuerPetruccione2002, RivasHuelga2012, Vacchini2024, Rivas2010, Manzano2020} for more details. Consider an open quantum system $S$ with Hilbert space $\MH_S$ which is in contact with a thermodynamic reservoir $R$ with Hilbert space $\MH_R$. The Hilbert space of the combined system is $\MH = \MH_S \otimes \MH_R$, and the total Hamiltonian may be decomposed as $\hat{H} = \hat{H}_S \otimes \id_R + \id_S \otimes \hat{H}_R + \alpha \hat{H}_\mathrm{int}$, where $\hat{H}_\mathrm{int}$ mediates the interaction between $S$ and $R$, and $\alpha \in \R$ determines the strength of the interaction. Let $\hat{\rho}$ denote the density operator of the total system. In the interaction picture, the von Neumann equation for $\hat{\rho}$ takes the following form \cite[Eq. (3.111)]{BreuerPetruccione2002}:
\begin{equation*}
    \frac{\diff \hat{\rho}(t)}{\diff t} = - \frac{\ii \alpha}{\hbar} \, \bigl[ \hat{H}_\mathrm{int}(t), \hat{\rho}(t) \bigr] \ .
\end{equation*}

Assuming a weak coupling regime ($\alpha \ll 1$), one can solve this equation in terms of a perturbative expansion, for example using the Nakajima-Zwanzig projection operator technique \cite{Nakajima1958, Zwanzig1960}, to obtain a master equation for the reduced state $\hat{\rho}_S (t) \ce \tr_R \bigl(\hat{\rho}(t)\bigr)$ of the open system, cf. \cite[Sections 3.3 and 9.1]{BreuerPetruccione2002}, \cite[Section 5.1]{RivasHuelga2012}, \cite[Sections 5.5 and 6.2]{Vacchini2024}. To this end, two additional approximations are implemented: (i) the \emph{Born approximation}, according to which the initial state of the total system factorizes as $\hat{\rho}(0) = \hat{\rho}_S(0) \otimes \hat{\rho}_R$ with $\hat{\rho}_R$ a canonical Gibbs state, and (ii) the \emph{Markov approximation}, which assumes that the typical timescale of the reservoir $\tau_R$ is small compared to the timescale $\tau_S$ over which $\hat{\rho}_S$ varies. Under these assumptions, one obtains \cite[Eq. (3.118)]{BreuerPetruccione2002}
\begin{equation}\label{eq:RedfieldIntegral}
    \frac{\diff \hat{\rho}_S(t)}{\diff t} = - \left(\frac{\alpha}{\hbar}\right)^2 \int_{0}^{+\infty} \tr_R \Bigl( \bigl[ \hat{H}_\mathrm{int}(t), [\hat{H}_\mathrm{int}(t - t^\prime), \hat{\rho}_S(t) \otimes \hat{\rho}_R] \bigr] \Bigr) \diff t^\prime + \MO(\alpha^3) \ .
\end{equation}
This equation is called \emph{Redfield equation} \cite{Redfield1957} or quantum master equation in the Born-Markov approximation. It has been well-known for a long time that solutions of \eqref{eq:RedfieldIntegral} need not be positive, i.e., they may possess negative eigenvalues; see Refs. \cite{Davies1974, Davies1976, Dümcke1979, Suarez1992, Benatti2005, Whitney2008, Tupkary2023, Campaioli2024} and references therein. In order to obtain a master equation that preserves positivity, one has to perform an additional approximation, the so-called \emph{secular} or \emph{rotating-wave approximation}, in which one averages over terms that oscillate over a timescale $\tau_0$ that is much larger than $\tau_R$ but much smaller than $\tau_S$. The resulting master equation takes the form \cite[Eq. (3.66)]{BreuerPetruccione2002}
\begin{equation}\label{eq:LindbladApp}
    \frac{\diff \hat{\rho}_S(t)}{\diff t} = -\frac{\ii}{\hbar} \, [\hat{H}, \hat{\rho}_S(t)] + \alpha^2 \sum_{j} \lambda_j \left(\hat{L}_j \hat{\rho}_S(t) \hat{L}_j^\dagger - \frac{1}{2} \, \bigl\{ \hat{L}_j^\dagger \hat{L}_j \hat{\rho}_S(t) \bigr\}\right) \ .
\end{equation}
This is the \emph{Lindblad equation} which was independently obtained for bounded Hamiltonians on infinite-dimensional spaces by Lindblad \cite{Lindblad1976}, and for finite-dimensional spaces by Gorini, Kossakowski, and Sudarshan \cite{Gorini1976}. It turns out that a necessary and sufficient condition for $\hat{\rho}_S(t)$ to be a positive operator for all $t > 0$ is that the master equation can be written in the form \eqref{eq:LindbladApp}; this can be formalized \cite[Theorem 4.2.1]{RivasHuelga2012}, \cite[Theorem 5.1]{Vacchini2024} by the following

\begin{theorem}
    Suppose that $\dim \MH_S < + \infty$. Then the time-evolution $\hat{\rho}_S(t) = \Phi(t) \hat{\rho}_S(0)$ is completely positive, meaning that $\hat{\rho}_S(t)$ is a positive operator for all $t > 0$, if and only if the generator $\ML$ of $(\Phi(t))_{t \ge 0}$, defined via $\frac{\diff}{\diff t} \, \hat{\rho}_S(t) = \ML \hat{\rho}_S(t)$, is given by
    \begin{equation*}
        \ML \hat{\rho}_S = -\frac{\ii}{\hbar} \, [\hat{H}, \hat{\rho}_S] + \alpha^2 \sum_{j} \lambda_j \left(\hat{L}_j \hat{\rho}_S \hat{L}_j^\dagger - \frac{1}{2} \, \bigl\{ \hat{L}_j^\dagger \hat{L}_j \hat{\rho}_S \bigr\}\right)
    \end{equation*}
    with $\lambda_j \ge 0$, $\hat{L}_j$ bounded operators on $\MH_S$, and $\hat{H}$ a self-adjoint operator on $\MH_S$.
\end{theorem}

\section{Limit to equilibrium in the NPI method}\label{app:limitEquilibrium}

In PIMD the calculation of equilibrium time correlation functions $C_{AB}(t)$ for two physical observables $\hat{A}$ and $\hat{B}$ has been a subject of high interest. These functions can be defined in different ways; while fully equivalent in a quantum-mechanical treatment, they are no longer equivalent if classical approximations are used; see, e.g., Refs. \cite{Jang1999, Habershon2013, Ciccotti2010} and references therein. We shall use the definition of $C_{AB}(t)$ from Ref. \cite[Eq. (1)]{Habershon2013} and show that in the limit of equilibrium (i.e., when the dissipative part of the Lindblad equation can be neglected), the NPI method consistently leads to the same formula which is obtained through the PIMD method only \cite[Eq. (11)]{Habershon2013}.

The time correlation function $C_{AB}(t)$ for two observables $\hat{A}$ and $\hat{B}$ in the Heisenberg representation is given by \cite[Eq. (1)]{Habershon2013}
\begin{equation}\label{eq:timeCorrFunc}
    C_{AB}(t) \ce \tr \bigl( \hat{\rho}(0) \hat{A}(0) \hat{B}(t) \bigr) \ .
\end{equation}
Defining $\hat{F}(t) \ce \hat{A}(0) \hat{B}(t)$, it follows that $C_{AB} (t) = \tr \bigl( \hat{\rho}(0) \hat{F}(t) \bigr) = \braket{\hat{F}}(t)$ is the time-dependent expectation value of $\hat{F}(t)$ from \cref{eq:expValTransform}. If we take the limit $\lambda_j \to 0$ in the Lindblad equation \eqref{eq:Lindblad}, that is, if we are in the limit of equilibrium, the results from \cref{sec:PIMD} can be applied straightforwardly; in particular, the expectation value $\braket{\hat{F}}(t)$ can be computed using \cref{eq:expValPIMD} which yields
\begin{align}\label{eq:timeCorrFuncPIMD}
    \begin{split}
       C_{AB}(t) = \lim_{P \to + \infty} \, \frac{1}{Z_P} &\int_\R \, \prod_{i=1}^{N} \diff^3 \bfp_i^{(1)} \dotsm \diff^3 \bfp_i^{(P)} \int_{D(L)} \, \prod_{i=1}^{N} \diff^3 \bfr_i^{(1)} \dotsm \diff^3 \bfr_i^{(P)} \\
            &\times \Bigl[A_P \bigl(\bfr(0)\bigr) B_P \bigl(\bfr(t)\bigr) \, \ee^{- \beta H_P (\bfr_0, \bfp_0)}\Bigr] \ ,
    \end{split}
\end{align}
where $A_P (\bfr) = \frac{1}{P} \sum_{k=1}^{P} A (\bfr_k)$, with $\bfr$ corresponding to a point on the trajectory at time $t = t_0$, and equivalently for $B_P (\bfr)$, with $\bfr$ corresponding to a point at time $t = t_0 + m \Delta t$, and finally $H_P$ is defined in \cref{eq:effectiveQuantumH}. The above expression is the equivalent of \cite[Eq. (11)]{Habershon2013} for an $N$-particle system, thus showing consistency.

While the formal correspondence between \cref{eq:timeCorrFunc,eq:expValTransform} is trivial, the important point to note is that the corresponding numerical algorithms, one based on the full equilibrium approach following \cref{eq:timeCorrFuncPIMD} from Ref. \cite{Habershon2013}, and the other employing the numerical procedure of D-NEMD in the limit of equilibrium, are equivalent. From the algorithmic point of view, the key difference is that in D-NEMD, $\hat{B}(t)$ is calculated along branched trajectories while in the other method, one simply produces a single equilibrium trajectory along which \cref{eq:timeCorrFuncPIMD} is calculated. Realizing that in the limit of equilibrium, each branched trajectory is equivalent to the equilibrium trajectory from a certain time on (see \cref{fig:limiteq} for an illustration), the calculation of $C_{AB} (t)$ following \cref{mdquantnoneq} with $\hat{F}(t) = \hat{A}(0) \hat{B}(t)$ reduces to an average over time intervals of equal length along the equilibrium trajectory. One then proceeds by building a time series with progressively larger time interval span, exactly as done in Refs. \cite{Agarwal2015, Agarwal2016, Agarwal2017} following \cref{eq:timeCorrFuncPIMD} and using a full equilibrium PIMD approach.

\begin{figure}
    \centering
    \scalebox{0.8}{%
    \begin{tikzpicture}[font=\large]

        \foreach \x/\y [count=\i from 0] in {0/0, 3/1, 6/0.25, 9/0.5, 12/-0.25, 15/0.25, 18/0}
        {
            \coordinate (t\i) at (\x,\y);
        }
    
        \foreach \i/\l in {0/0, 1/1, 2/2, 5/n-1, 6/n}
        {
            \node [circle, fill, inner sep=2pt] at (t\i) {};
            \node [above, yshift=0.1cm] at (t\i) {$t_{\l}$};
        }
    
        \draw [very thick] (t0) .. controls (1,1) and (2,1.25) .. (t1)
                                .. controls (4,0.75) and (5,0.25) .. (t2)
                                .. controls (7,0.25) and (8,0.5) .. (t3)
                                .. controls (10,0.5) and (11,-0.25) .. (t4)
                                .. controls (13,-0.25) and (14,0.25) .. (t5)
                                .. controls (16,0.25) and (17,0) .. (t6);

        \foreach \i/\x in {0/0, 1/3, 2/6, 5/15}
        {
            \foreach \j/\y in {1/-2, 2/-4, 3/-6, 4/-8, 5/-10}
            {
                \coordinate (t\i\j) at (\x,\y);
                \ifthenelse{\inteval{\i + \j} < 4}
                {
                    \pgfmathsetmacro{\index}{\i + \j}
                    \node [circle, fill, inner sep=2pt] at (t\i\j) {};
                    \node [above right, yshift=0.1cm] at (t\i\j) {$t_{\pgfmathprintnumber{\index}}^\prime \equiv t_{\pgfmathprintnumber{\index}}$};
                }
                {}
                \ifthenelse{\inteval{\i + \j} = 4 \OR \(\i = 5 \AND \j = 1\)}
                {
                    \node [circle, fill, inner sep=2pt] at (t\i\j) {};
                    \node [above right, yshift=0.1cm] at (t\i\j) {$t_n^\prime \equiv t_n$};
                }
                {}
            }
        }
    
        \foreach \i in {0, 1, 2, 5}
        {
            \draw [trajectory] (t\i) -- (t\i1);
            \foreach \j [count=\n] in {2, 3, 4}
            {
                \ifthenelse{\inteval{\i + \j} < 4}
                {
                    \draw [trajectory] (t\i\n) -- (t\i\j);
                }
                {}
                \ifthenelse{\inteval{\i + \j} = 4}
                {
                    \draw [trajectory, dashed] (t\i\n) -- (t\i\j);
                }
                {}
            }
        }
    
        \foreach \j in {1, 2, 3}
        {
            \foreach \i/\f in {0/1, 1/2, 2/5}
            {
                \ifthenelse{\inteval{\j + \i} < 3}
                {
                    \draw [trajectory] (t\i\j) -- (t\f\j);
                }
                {}
                \ifthenelse{\inteval{\j + \i} = 3}
                {
                    \draw [trajectory, dashed] (t\i\j) -- (t\f\j);
                }
                {}
            }
        }
    
        \foreach \x/\y in {10.5/-2.5, 4.5/-4.5, 1.5/-6.5}
        {
            \node at (\x,\y) {$\hdots$};
        }
    
        \foreach \x/\y in {-0.5/-7, 2.5/-5, 5.5/-3}
        {
            \node at (\x,\y) {$\vdots$};
        }
    
        \node at (9,-9) {$\vert t_i - t_{i+1} \vert = \Delta t = \vert t_i - t_{i+1}^\prime \vert \quad (i \in \{0, 1, \dotsc, n - 1\})$};
    
    \end{tikzpicture}}
    \caption{Illustration of the limit to equilibrium of the D-NEMD method. The same explanations as given in \cref{fig:d-nemd} apply, but now the vertical lines correspond to trajectories that are equivalent to the equilibrium trajectory starting from the point taken from the horizontal curve, which is the main equilibrium trajectory.}
    \label{fig:limiteq}
\end{figure}

\section{Technical details of the simulation}\label{app:techdet}

All simulations were performed using the LAMMPS molecular dynamics package \cite{plimpton1995fast, thompson2022lammps}. Water molecules were modeled using the flexible q-TIP4P/F force field \cite{habershon2009competing}, which combines Lennard--Jones interactions between oxygen atoms with long-range electrostatics and flexible intramolecular degrees of freedom. The intramolecular O--H bonds were described by a Morse potential and the intramolecular H--O--H angle by a harmonic potential. Long-range Coulomb interactions were treated using the PPPM method adapted for TIP4P-type models with a real-space cutoff of 9~\AA. Quantum effects were included via path-integral molecular dynamics, in which each atom is represented by a polymer ring of $P$ beads. Simulations were carried out with $P = 1, 16, 32, 64$, corresponding to a classical ($P=1$) and different levels of a quantum description of the atoms ($P=16, 32, 64$). Time integration for the equilibrium state was performed using the BAOAB integrator combined with a PILE-L Langevin thermostat at 300~K with a relaxation time of 1~ps, employing the normal-mode PIMD technique \cite{ceriotti2010efficient}. The system in equilibrium was simulated under periodic boundary conditions in all directions using a timestep of 0.5~fs for a total of 0.2~ns.

After equilibration, non-equilibrium molecular dynamics simulations were performed according to our proposed model along 27 different branches, each starting at a point of the equilibrium trajectory and lasting 5~ns, by imposing a thermal gradient along the $z$-direction following established protocols \cite{fernando2017non}. To maintain periodic boundary conditions along the $z$-direction, the simulation domain was divided into a symmetric hot-middle-cold-middle-hot arrangement, with the hot regions chosen to be half the spatial extent of the remaining regions \cite{fernando2017non}. The hot regions were maintained at 330~K, while the cold region was maintained at 280~K, with the middle regions reaching a steady state of uniform temperature of about 305~K as expected, departing from the initial non-uniform $z$-dependent temperature along the chain. The hot and cold regions where each coupled to a (region-specific) Langevin thermostat, assuring the corresponding desired temperature \cite{Tuckerman2023}.

The overall system contains a total of 350 molecules, 87 for each middle region which corresponds to the system under observation, and the rest equally distributed among the reservoirs which were strongly thermostated at different temperatures when the non equilibrium process was initiated. The temperature gradients and heat fluxes were computed as time averages over production runs of (the last) 2.5~ns for each of the 27 non-equilibrium branched trajectories. For PIMD simulations, all observables were further averaged over the $P$ beads to obtain physically meaningful estimates. In particular, the heat flux through atoms in the middle region was calculated using the microscopic Irving--Kirkwood expression, which combines per-atom kinetic energy, potential energy, and virial stress contributions \cite{surblys2021methodology}, and was averaged over both time and beads. Local temperature profiles were obtained by spatially binning atoms along the $z$-direction and time-averaging over the production runs.

\bibliography{lind_revision.bib}

\end{document}

%% file: figures/thermal_profiles.pgf
\begingroup%
\makeatletter%
\begin{pgfpicture}%
\pgfpathrectangle{\pgfpointorigin}{\pgfqpoint{9.659354in}{4.700785in}}%
\pgfusepath{use as bounding box, clip}%
\begin{pgfscope}%
\pgfsetbuttcap%
\pgfsetmiterjoin%
\definecolor{currentfill}{rgb}{1.000000,1.000000,1.000000}%
\pgfsetfillcolor{currentfill}%
\pgfsetlinewidth{0.000000pt}%
\definecolor{currentstroke}{rgb}{1.000000,1.000000,1.000000}%
\pgfsetstrokecolor{currentstroke}%
\pgfsetdash{}{0pt}%
\pgfpathmoveto{\pgfqpoint{0.000000in}{0.000000in}}%
\pgfpathlineto{\pgfqpoint{9.659354in}{0.000000in}}%
\pgfpathlineto{\pgfqpoint{9.659354in}{4.700785in}}%
\pgfpathlineto{\pgfqpoint{0.000000in}{4.700785in}}%
\pgfpathlineto{\pgfqpoint{0.000000in}{0.000000in}}%
\pgfpathclose%
\pgfusepath{fill}%
\end{pgfscope}%
\begin{pgfscope}%
\pgfsetbuttcap%
\pgfsetmiterjoin%
\definecolor{currentfill}{rgb}{1.000000,1.000000,1.000000}%
\pgfsetfillcolor{currentfill}%
\pgfsetlinewidth{0.000000pt}%
\definecolor{currentstroke}{rgb}{0.000000,0.000000,0.000000}%
\pgfsetstrokecolor{currentstroke}%
\pgfsetstrokeopacity{0.000000}%
\pgfsetdash{}{0pt}%
\pgfpathmoveto{\pgfqpoint{0.960541in}{0.793563in}}%
\pgfpathlineto{\pgfqpoint{9.497451in}{0.793563in}}%
\pgfpathlineto{\pgfqpoint{9.497451in}{4.600785in}}%
\pgfpathlineto{\pgfqpoint{0.960541in}{4.600785in}}%
\pgfpathlineto{\pgfqpoint{0.960541in}{0.793563in}}%
\pgfpathclose%
\pgfusepath{fill}%
\end{pgfscope}%
\begin{pgfscope}%
\pgfpathrectangle{\pgfqpoint{0.960541in}{0.793563in}}{\pgfqpoint{8.536910in}{3.807222in}}%
\pgfusepath{clip}%
\pgfsetbuttcap%
\pgfsetmiterjoin%
\definecolor{currentfill}{rgb}{1.000000,0.490196,0.490196}%
\pgfsetfillcolor{currentfill}%
\pgfsetfillopacity{0.400000}%
\pgfsetlinewidth{1.003750pt}%
\definecolor{currentstroke}{rgb}{1.000000,0.490196,0.490196}%
\pgfsetstrokecolor{currentstroke}%
\pgfsetstrokeopacity{0.400000}%
\pgfsetdash{}{0pt}%
\pgfpathmoveto{\pgfqpoint{0.960541in}{0.793563in}}%
\pgfpathlineto{\pgfqpoint{1.302018in}{0.793563in}}%
\pgfpathlineto{\pgfqpoint{1.302018in}{4.600785in}}%
\pgfpathlineto{\pgfqpoint{0.960541in}{4.600785in}}%
\pgfpathlineto{\pgfqpoint{0.960541in}{0.793563in}}%
\pgfpathclose%
\pgfusepath{stroke,fill}%
\end{pgfscope}%
\begin{pgfscope}%
\pgfpathrectangle{\pgfqpoint{0.960541in}{0.793563in}}{\pgfqpoint{8.536910in}{3.807222in}}%
\pgfusepath{clip}%
\pgfsetbuttcap%
\pgfsetmiterjoin%
\definecolor{currentfill}{rgb}{0.556863,0.556863,1.000000}%
\pgfsetfillcolor{currentfill}%
\pgfsetfillopacity{0.400000}%
\pgfsetlinewidth{1.003750pt}%
\definecolor{currentstroke}{rgb}{0.556863,0.556863,1.000000}%
\pgfsetstrokecolor{currentstroke}%
\pgfsetstrokeopacity{0.400000}%
\pgfsetdash{}{0pt}%
\pgfpathmoveto{\pgfqpoint{9.155975in}{0.793563in}}%
\pgfpathlineto{\pgfqpoint{9.838927in}{0.793563in}}%
\pgfpathlineto{\pgfqpoint{9.838927in}{4.600785in}}%
\pgfpathlineto{\pgfqpoint{9.155975in}{4.600785in}}%
\pgfpathlineto{\pgfqpoint{9.155975in}{0.793563in}}%
\pgfpathclose%
\pgfusepath{stroke,fill}%
\end{pgfscope}%
\begin{pgfscope}%
\pgfsetbuttcap%
\pgfsetroundjoin%
\definecolor{currentfill}{rgb}{0.000000,0.000000,0.000000}%
\pgfsetfillcolor{currentfill}%
\pgfsetlinewidth{2.007500pt}%
\definecolor{currentstroke}{rgb}{0.000000,0.000000,0.000000}%
\pgfsetstrokecolor{currentstroke}%
\pgfsetdash{}{0pt}%
\pgfsys@defobject{currentmarker}{\pgfqpoint{0.000000in}{0.000000in}}{\pgfqpoint{0.000000in}{0.083333in}}{%
\pgfpathmoveto{\pgfqpoint{0.000000in}{0.000000in}}%
\pgfpathlineto{\pgfqpoint{0.000000in}{0.083333in}}%
\pgfusepath{stroke,fill}%
}%
\begin{pgfscope}%
\pgfsys@transformshift{0.960541in}{0.793563in}%
\pgfsys@useobject{currentmarker}{}%
\end{pgfscope}%
\end{pgfscope}%
\begin{pgfscope}%
\definecolor{textcolor}{rgb}{0.000000,0.000000,0.000000}%
\pgfsetstrokecolor{textcolor}%
\pgfsetfillcolor{textcolor}%
\pgftext[x=0.960541in,y=0.738007in,,top]{\color{textcolor}{\sffamily\fontsize{16.000000}{19.200000}\selectfont\catcode`\^=\active\def^{\ifmmode\sp\else\^{}\fi}\catcode`\%=\active\def
\end{pgfscope}%
\begin{pgfscope}%
\pgfsetbuttcap%
\pgfsetroundjoin%
\definecolor{currentfill}{rgb}{0.000000,0.000000,0.000000}%
\pgfsetfillcolor{currentfill}%
\pgfsetlinewidth{2.007500pt}%
\definecolor{currentstroke}{rgb}{0.000000,0.000000,0.000000}%
\pgfsetstrokecolor{currentstroke}%
\pgfsetdash{}{0pt}%
\pgfsys@defobject{currentmarker}{\pgfqpoint{0.000000in}{0.000000in}}{\pgfqpoint{0.000000in}{0.083333in}}{%
\pgfpathmoveto{\pgfqpoint{0.000000in}{0.000000in}}%
\pgfpathlineto{\pgfqpoint{0.000000in}{0.083333in}}%
\pgfusepath{stroke,fill}%
}%
\begin{pgfscope}%
\pgfsys@transformshift{5.228996in}{0.793563in}%
\pgfsys@useobject{currentmarker}{}%
\end{pgfscope}%
\end{pgfscope}%
\begin{pgfscope}%
\definecolor{textcolor}{rgb}{0.000000,0.000000,0.000000}%
\pgfsetstrokecolor{textcolor}%
\pgfsetfillcolor{textcolor}%
\pgftext[x=5.228996in,y=0.738007in,,top]{\color{textcolor}{\sffamily\fontsize{16.000000}{19.200000}\selectfont\catcode`\^=\active\def^{\ifmmode\sp\else\^{}\fi}\catcode`\%=\active\def
\end{pgfscope}%
\begin{pgfscope}%
\pgfsetbuttcap%
\pgfsetroundjoin%
\definecolor{currentfill}{rgb}{0.000000,0.000000,0.000000}%
\pgfsetfillcolor{currentfill}%
\pgfsetlinewidth{2.007500pt}%
\definecolor{currentstroke}{rgb}{0.000000,0.000000,0.000000}%
\pgfsetstrokecolor{currentstroke}%
\pgfsetdash{}{0pt}%
\pgfsys@defobject{currentmarker}{\pgfqpoint{0.000000in}{0.000000in}}{\pgfqpoint{0.000000in}{0.083333in}}{%
\pgfpathmoveto{\pgfqpoint{0.000000in}{0.000000in}}%
\pgfpathlineto{\pgfqpoint{0.000000in}{0.083333in}}%
\pgfusepath{stroke,fill}%
}%
\begin{pgfscope}%
\pgfsys@transformshift{9.497451in}{0.793563in}%
\pgfsys@useobject{currentmarker}{}%
\end{pgfscope}%
\end{pgfscope}%
\begin{pgfscope}%
\definecolor{textcolor}{rgb}{0.000000,0.000000,0.000000}%
\pgfsetstrokecolor{textcolor}%
\pgfsetfillcolor{textcolor}%
\pgftext[x=9.497451in,y=0.738007in,,top]{\color{textcolor}{\sffamily\fontsize{16.000000}{19.200000}\selectfont\catcode`\^=\active\def^{\ifmmode\sp\else\^{}\fi}\catcode`\%=\active\def
\end{pgfscope}%
\begin{pgfscope}%
\definecolor{textcolor}{rgb}{0.000000,0.000000,0.000000}%
\pgfsetstrokecolor{textcolor}%
\pgfsetfillcolor{textcolor}%
\pgftext[x=5.228996in,y=0.384057in,,top]{\color{textcolor}{\sffamily\fontsize{18.000000}{21.600000}\selectfont\catcode`\^=\active\def^{\ifmmode\sp\else\^{}\fi}\catcode`\%=\active\def
\end{pgfscope}%
\begin{pgfscope}%
\pgfsetbuttcap%
\pgfsetroundjoin%
\definecolor{currentfill}{rgb}{0.000000,0.000000,0.000000}%
\pgfsetfillcolor{currentfill}%
\pgfsetlinewidth{2.007500pt}%
\definecolor{currentstroke}{rgb}{0.000000,0.000000,0.000000}%
\pgfsetstrokecolor{currentstroke}%
\pgfsetdash{}{0pt}%
\pgfsys@defobject{currentmarker}{\pgfqpoint{0.000000in}{0.000000in}}{\pgfqpoint{0.083333in}{0.000000in}}{%
\pgfpathmoveto{\pgfqpoint{0.000000in}{0.000000in}}%
\pgfpathlineto{\pgfqpoint{0.083333in}{0.000000in}}%
\pgfusepath{stroke,fill}%
}%
\begin{pgfscope}%
\pgfsys@transformshift{0.960541in}{1.112961in}%
\pgfsys@useobject{currentmarker}{}%
\end{pgfscope}%
\end{pgfscope}%
\begin{pgfscope}%
\definecolor{textcolor}{rgb}{0.000000,0.000000,0.000000}%
\pgfsetstrokecolor{textcolor}%
\pgfsetfillcolor{textcolor}%
\pgftext[x=0.480832in, y=1.028543in, left, base]{\color{textcolor}{\sffamily\fontsize{16.000000}{19.200000}\selectfont\catcode`\^=\active\def^{\ifmmode\sp\else\^{}\fi}\catcode`\%=\active\def
\end{pgfscope}%
\begin{pgfscope}%
\pgfsetbuttcap%
\pgfsetroundjoin%
\definecolor{currentfill}{rgb}{0.000000,0.000000,0.000000}%
\pgfsetfillcolor{currentfill}%
\pgfsetlinewidth{2.007500pt}%
\definecolor{currentstroke}{rgb}{0.000000,0.000000,0.000000}%
\pgfsetstrokecolor{currentstroke}%
\pgfsetdash{}{0pt}%
\pgfsys@defobject{currentmarker}{\pgfqpoint{0.000000in}{0.000000in}}{\pgfqpoint{0.083333in}{0.000000in}}{%
\pgfpathmoveto{\pgfqpoint{0.000000in}{0.000000in}}%
\pgfpathlineto{\pgfqpoint{0.083333in}{0.000000in}}%
\pgfusepath{stroke,fill}%
}%
\begin{pgfscope}%
\pgfsys@transformshift{0.960541in}{1.762936in}%
\pgfsys@useobject{currentmarker}{}%
\end{pgfscope}%
\end{pgfscope}%
\begin{pgfscope}%
\definecolor{textcolor}{rgb}{0.000000,0.000000,0.000000}%
\pgfsetstrokecolor{textcolor}%
\pgfsetfillcolor{textcolor}%
\pgftext[x=0.480832in, y=1.678517in, left, base]{\color{textcolor}{\sffamily\fontsize{16.000000}{19.200000}\selectfont\catcode`\^=\active\def^{\ifmmode\sp\else\^{}\fi}\catcode`\%=\active\def
\end{pgfscope}%
\begin{pgfscope}%
\pgfsetbuttcap%
\pgfsetroundjoin%
\definecolor{currentfill}{rgb}{0.000000,0.000000,0.000000}%
\pgfsetfillcolor{currentfill}%
\pgfsetlinewidth{2.007500pt}%
\definecolor{currentstroke}{rgb}{0.000000,0.000000,0.000000}%
\pgfsetstrokecolor{currentstroke}%
\pgfsetdash{}{0pt}%
\pgfsys@defobject{currentmarker}{\pgfqpoint{0.000000in}{0.000000in}}{\pgfqpoint{0.083333in}{0.000000in}}{%
\pgfpathmoveto{\pgfqpoint{0.000000in}{0.000000in}}%
\pgfpathlineto{\pgfqpoint{0.083333in}{0.000000in}}%
\pgfusepath{stroke,fill}%
}%
\begin{pgfscope}%
\pgfsys@transformshift{0.960541in}{2.412910in}%
\pgfsys@useobject{currentmarker}{}%
\end{pgfscope}%
\end{pgfscope}%
\begin{pgfscope}%
\definecolor{textcolor}{rgb}{0.000000,0.000000,0.000000}%
\pgfsetstrokecolor{textcolor}%
\pgfsetfillcolor{textcolor}%
\pgftext[x=0.480832in, y=2.328492in, left, base]{\color{textcolor}{\sffamily\fontsize{16.000000}{19.200000}\selectfont\catcode`\^=\active\def^{\ifmmode\sp\else\^{}\fi}\catcode`\%=\active\def
\end{pgfscope}%
\begin{pgfscope}%
\pgfsetbuttcap%
\pgfsetroundjoin%
\definecolor{currentfill}{rgb}{0.000000,0.000000,0.000000}%
\pgfsetfillcolor{currentfill}%
\pgfsetlinewidth{2.007500pt}%
\definecolor{currentstroke}{rgb}{0.000000,0.000000,0.000000}%
\pgfsetstrokecolor{currentstroke}%
\pgfsetdash{}{0pt}%
\pgfsys@defobject{currentmarker}{\pgfqpoint{0.000000in}{0.000000in}}{\pgfqpoint{0.083333in}{0.000000in}}{%
\pgfpathmoveto{\pgfqpoint{0.000000in}{0.000000in}}%
\pgfpathlineto{\pgfqpoint{0.083333in}{0.000000in}}%
\pgfusepath{stroke,fill}%
}%
\begin{pgfscope}%
\pgfsys@transformshift{0.960541in}{3.062885in}%
\pgfsys@useobject{currentmarker}{}%
\end{pgfscope}%
\end{pgfscope}%
\begin{pgfscope}%
\definecolor{textcolor}{rgb}{0.000000,0.000000,0.000000}%
\pgfsetstrokecolor{textcolor}%
\pgfsetfillcolor{textcolor}%
\pgftext[x=0.480832in, y=2.978466in, left, base]{\color{textcolor}{\sffamily\fontsize{16.000000}{19.200000}\selectfont\catcode`\^=\active\def^{\ifmmode\sp\else\^{}\fi}\catcode`\%=\active\def
\end{pgfscope}%
\begin{pgfscope}%
\pgfsetbuttcap%
\pgfsetroundjoin%
\definecolor{currentfill}{rgb}{0.000000,0.000000,0.000000}%
\pgfsetfillcolor{currentfill}%
\pgfsetlinewidth{2.007500pt}%
\definecolor{currentstroke}{rgb}{0.000000,0.000000,0.000000}%
\pgfsetstrokecolor{currentstroke}%
\pgfsetdash{}{0pt}%
\pgfsys@defobject{currentmarker}{\pgfqpoint{0.000000in}{0.000000in}}{\pgfqpoint{0.083333in}{0.000000in}}{%
\pgfpathmoveto{\pgfqpoint{0.000000in}{0.000000in}}%
\pgfpathlineto{\pgfqpoint{0.083333in}{0.000000in}}%
\pgfusepath{stroke,fill}%
}%
\begin{pgfscope}%
\pgfsys@transformshift{0.960541in}{3.712859in}%
\pgfsys@useobject{currentmarker}{}%
\end{pgfscope}%
\end{pgfscope}%
\begin{pgfscope}%
\definecolor{textcolor}{rgb}{0.000000,0.000000,0.000000}%
\pgfsetstrokecolor{textcolor}%
\pgfsetfillcolor{textcolor}%
\pgftext[x=0.480832in, y=3.628441in, left, base]{\color{textcolor}{\sffamily\fontsize{16.000000}{19.200000}\selectfont\catcode`\^=\active\def^{\ifmmode\sp\else\^{}\fi}\catcode`\%=\active\def
\end{pgfscope}%
\begin{pgfscope}%
\pgfsetbuttcap%
\pgfsetroundjoin%
\definecolor{currentfill}{rgb}{0.000000,0.000000,0.000000}%
\pgfsetfillcolor{currentfill}%
\pgfsetlinewidth{2.007500pt}%
\definecolor{currentstroke}{rgb}{0.000000,0.000000,0.000000}%
\pgfsetstrokecolor{currentstroke}%
\pgfsetdash{}{0pt}%
\pgfsys@defobject{currentmarker}{\pgfqpoint{0.000000in}{0.000000in}}{\pgfqpoint{0.083333in}{0.000000in}}{%
\pgfpathmoveto{\pgfqpoint{0.000000in}{0.000000in}}%
\pgfpathlineto{\pgfqpoint{0.083333in}{0.000000in}}%
\pgfusepath{stroke,fill}%
}%
\begin{pgfscope}%
\pgfsys@transformshift{0.960541in}{4.362834in}%
\pgfsys@useobject{currentmarker}{}%
\end{pgfscope}%
\end{pgfscope}%
\begin{pgfscope}%
\definecolor{textcolor}{rgb}{0.000000,0.000000,0.000000}%
\pgfsetstrokecolor{textcolor}%
\pgfsetfillcolor{textcolor}%
\pgftext[x=0.480832in, y=4.278416in, left, base]{\color{textcolor}{\sffamily\fontsize{16.000000}{19.200000}\selectfont\catcode`\^=\active\def^{\ifmmode\sp\else\^{}\fi}\catcode`\%=\active\def
\end{pgfscope}%
\begin{pgfscope}%
\definecolor{textcolor}{rgb}{0.000000,0.000000,0.000000}%
\pgfsetstrokecolor{textcolor}%
\pgfsetfillcolor{textcolor}%
\pgftext[x=0.341943in,y=2.697174in,,bottom,rotate=90.000000]{\color{textcolor}{\sffamily\fontsize{18.000000}{21.600000}\selectfont\catcode`\^=\active\def^{\ifmmode\sp\else\^{}\fi}\catcode`\%=\active\def
\end{pgfscope}%
\begin{pgfscope}%
\pgfpathrectangle{\pgfqpoint{0.960541in}{0.793563in}}{\pgfqpoint{8.536910in}{3.807222in}}%
\pgfusepath{clip}%
\pgfsetrectcap%
\pgfsetroundjoin%
\pgfsetlinewidth{3.011250pt}%
\definecolor{currentstroke}{rgb}{0.000000,0.000000,1.000000}%
\pgfsetstrokecolor{currentstroke}%
\pgfsetdash{}{0pt}%
\pgfpathmoveto{\pgfqpoint{0.960541in}{4.413411in}}%
\pgfpathlineto{\pgfqpoint{1.045910in}{4.427729in}}%
\pgfpathlineto{\pgfqpoint{1.131279in}{4.356286in}}%
\pgfpathlineto{\pgfqpoint{1.216649in}{4.359915in}}%
\pgfpathlineto{\pgfqpoint{1.302018in}{4.345597in}}%
\pgfpathlineto{\pgfqpoint{1.336677in}{2.621931in}}%
\pgfpathlineto{\pgfqpoint{1.609859in}{2.861257in}}%
\pgfpathlineto{\pgfqpoint{1.883040in}{2.702491in}}%
\pgfpathlineto{\pgfqpoint{2.156221in}{2.737946in}}%
\pgfpathlineto{\pgfqpoint{2.429402in}{2.711006in}}%
\pgfpathlineto{\pgfqpoint{2.702583in}{2.717619in}}%
\pgfpathlineto{\pgfqpoint{2.975764in}{2.657700in}}%
\pgfpathlineto{\pgfqpoint{3.248945in}{2.659363in}}%
\pgfpathlineto{\pgfqpoint{3.522126in}{2.842991in}}%
\pgfpathlineto{\pgfqpoint{3.795307in}{2.766651in}}%
\pgfpathlineto{\pgfqpoint{4.068489in}{2.768692in}}%
\pgfpathlineto{\pgfqpoint{4.341670in}{2.816321in}}%
\pgfpathlineto{\pgfqpoint{4.614851in}{2.758961in}}%
\pgfpathlineto{\pgfqpoint{4.888032in}{2.671526in}}%
\pgfpathlineto{\pgfqpoint{5.161213in}{2.842928in}}%
\pgfpathlineto{\pgfqpoint{5.434394in}{2.812764in}}%
\pgfpathlineto{\pgfqpoint{5.707575in}{2.768973in}}%
\pgfpathlineto{\pgfqpoint{5.980756in}{2.802363in}}%
\pgfpathlineto{\pgfqpoint{6.253937in}{2.715218in}}%
\pgfpathlineto{\pgfqpoint{6.527119in}{2.705484in}}%
\pgfpathlineto{\pgfqpoint{6.800300in}{2.756253in}}%
\pgfpathlineto{\pgfqpoint{7.073481in}{2.878576in}}%
\pgfpathlineto{\pgfqpoint{7.346662in}{2.613978in}}%
\pgfpathlineto{\pgfqpoint{7.619843in}{2.749803in}}%
\pgfpathlineto{\pgfqpoint{7.893024in}{2.576471in}}%
\pgfpathlineto{\pgfqpoint{8.166205in}{2.782221in}}%
\pgfpathlineto{\pgfqpoint{8.439386in}{2.746900in}}%
\pgfpathlineto{\pgfqpoint{8.712567in}{2.666443in}}%
\pgfpathlineto{\pgfqpoint{8.985749in}{2.722552in}}%
\pgfpathlineto{\pgfqpoint{9.258930in}{1.436844in}}%
\pgfpathlineto{\pgfqpoint{9.326713in}{1.098502in}}%
\pgfpathlineto{\pgfqpoint{9.454766in}{1.080603in}}%
\pgfpathlineto{\pgfqpoint{9.499118in}{1.041124in}}%
\pgfusepath{stroke}%
\end{pgfscope}%
\begin{pgfscope}%
\pgfpathrectangle{\pgfqpoint{0.960541in}{0.793563in}}{\pgfqpoint{8.536910in}{3.807222in}}%
\pgfusepath{clip}%
\pgfsetrectcap%
\pgfsetroundjoin%
\pgfsetlinewidth{3.011250pt}%
\definecolor{currentstroke}{rgb}{0.000000,0.501961,0.000000}%
\pgfsetstrokecolor{currentstroke}%
\pgfsetdash{}{0pt}%
\pgfpathmoveto{\pgfqpoint{0.960541in}{4.413411in}}%
\pgfpathlineto{\pgfqpoint{1.045910in}{4.427729in}}%
\pgfpathlineto{\pgfqpoint{1.131279in}{4.356286in}}%
\pgfpathlineto{\pgfqpoint{1.216649in}{4.359915in}}%
\pgfpathlineto{\pgfqpoint{1.302018in}{4.345597in}}%
\pgfpathlineto{\pgfqpoint{1.336677in}{2.705086in}}%
\pgfpathlineto{\pgfqpoint{1.609859in}{2.701029in}}%
\pgfpathlineto{\pgfqpoint{1.883040in}{2.708532in}}%
\pgfpathlineto{\pgfqpoint{2.156221in}{2.741739in}}%
\pgfpathlineto{\pgfqpoint{2.429402in}{2.695467in}}%
\pgfpathlineto{\pgfqpoint{2.702583in}{2.715507in}}%
\pgfpathlineto{\pgfqpoint{2.975764in}{2.723727in}}%
\pgfpathlineto{\pgfqpoint{3.248945in}{2.701585in}}%
\pgfpathlineto{\pgfqpoint{3.522126in}{2.724068in}}%
\pgfpathlineto{\pgfqpoint{3.795307in}{2.707936in}}%
\pgfpathlineto{\pgfqpoint{4.068489in}{2.725879in}}%
\pgfpathlineto{\pgfqpoint{4.341670in}{2.697469in}}%
\pgfpathlineto{\pgfqpoint{4.614851in}{2.704312in}}%
\pgfpathlineto{\pgfqpoint{4.888032in}{2.712170in}}%
\pgfpathlineto{\pgfqpoint{5.161213in}{2.726652in}}%
\pgfpathlineto{\pgfqpoint{5.434394in}{2.698725in}}%
\pgfpathlineto{\pgfqpoint{5.707575in}{2.704568in}}%
\pgfpathlineto{\pgfqpoint{5.980756in}{2.728150in}}%
\pgfpathlineto{\pgfqpoint{6.253937in}{2.675535in}}%
\pgfpathlineto{\pgfqpoint{6.527119in}{2.716583in}}%
\pgfpathlineto{\pgfqpoint{6.800300in}{2.740204in}}%
\pgfpathlineto{\pgfqpoint{7.073481in}{2.725690in}}%
\pgfpathlineto{\pgfqpoint{7.346662in}{2.704648in}}%
\pgfpathlineto{\pgfqpoint{7.619843in}{2.735882in}}%
\pgfpathlineto{\pgfqpoint{7.893024in}{2.772982in}}%
\pgfpathlineto{\pgfqpoint{8.166205in}{2.714150in}}%
\pgfpathlineto{\pgfqpoint{8.439386in}{2.721091in}}%
\pgfpathlineto{\pgfqpoint{8.712567in}{2.738216in}}%
\pgfpathlineto{\pgfqpoint{8.985749in}{2.717286in}}%
\pgfpathlineto{\pgfqpoint{9.258930in}{1.091667in}}%
\pgfpathlineto{\pgfqpoint{9.326713in}{1.098502in}}%
\pgfpathlineto{\pgfqpoint{9.454766in}{1.080603in}}%
\pgfpathlineto{\pgfqpoint{9.499118in}{1.041124in}}%
\pgfusepath{stroke}%
\end{pgfscope}%
\begin{pgfscope}%
\pgfpathrectangle{\pgfqpoint{0.960541in}{0.793563in}}{\pgfqpoint{8.536910in}{3.807222in}}%
\pgfusepath{clip}%
\pgfsetrectcap%
\pgfsetroundjoin%
\pgfsetlinewidth{3.011250pt}%
\definecolor{currentstroke}{rgb}{1.000000,0.647059,0.000000}%
\pgfsetstrokecolor{currentstroke}%
\pgfsetdash{}{0pt}%
\pgfpathmoveto{\pgfqpoint{0.960541in}{4.413411in}}%
\pgfpathlineto{\pgfqpoint{1.045910in}{4.427729in}}%
\pgfpathlineto{\pgfqpoint{1.131279in}{4.356286in}}%
\pgfpathlineto{\pgfqpoint{1.216649in}{4.359915in}}%
\pgfpathlineto{\pgfqpoint{1.302018in}{4.345597in}}%
\pgfpathlineto{\pgfqpoint{1.336677in}{2.722485in}}%
\pgfpathlineto{\pgfqpoint{1.609859in}{2.722978in}}%
\pgfpathlineto{\pgfqpoint{1.883040in}{2.745138in}}%
\pgfpathlineto{\pgfqpoint{2.156221in}{2.712640in}}%
\pgfpathlineto{\pgfqpoint{2.429402in}{2.728957in}}%
\pgfpathlineto{\pgfqpoint{2.702583in}{2.688277in}}%
\pgfpathlineto{\pgfqpoint{2.975764in}{2.726138in}}%
\pgfpathlineto{\pgfqpoint{3.248945in}{2.699780in}}%
\pgfpathlineto{\pgfqpoint{3.522126in}{2.726778in}}%
\pgfpathlineto{\pgfqpoint{3.795307in}{2.719798in}}%
\pgfpathlineto{\pgfqpoint{4.068489in}{2.700933in}}%
\pgfpathlineto{\pgfqpoint{4.341670in}{2.738805in}}%
\pgfpathlineto{\pgfqpoint{4.614851in}{2.687552in}}%
\pgfpathlineto{\pgfqpoint{4.888032in}{2.734674in}}%
\pgfpathlineto{\pgfqpoint{5.161213in}{2.743312in}}%
\pgfpathlineto{\pgfqpoint{5.434394in}{2.710031in}}%
\pgfpathlineto{\pgfqpoint{5.707575in}{2.711941in}}%
\pgfpathlineto{\pgfqpoint{5.980756in}{2.721565in}}%
\pgfpathlineto{\pgfqpoint{6.253937in}{2.708207in}}%
\pgfpathlineto{\pgfqpoint{6.527119in}{2.727347in}}%
\pgfpathlineto{\pgfqpoint{6.800300in}{2.723441in}}%
\pgfpathlineto{\pgfqpoint{7.073481in}{2.718267in}}%
\pgfpathlineto{\pgfqpoint{7.346662in}{2.725585in}}%
\pgfpathlineto{\pgfqpoint{7.619843in}{2.699798in}}%
\pgfpathlineto{\pgfqpoint{7.893024in}{2.724612in}}%
\pgfpathlineto{\pgfqpoint{8.166205in}{2.692785in}}%
\pgfpathlineto{\pgfqpoint{8.439386in}{2.715630in}}%
\pgfpathlineto{\pgfqpoint{8.712567in}{2.710920in}}%
\pgfpathlineto{\pgfqpoint{8.985749in}{2.739835in}}%
\pgfpathlineto{\pgfqpoint{9.258930in}{1.094762in}}%
\pgfpathlineto{\pgfqpoint{9.326713in}{1.098502in}}%
\pgfpathlineto{\pgfqpoint{9.454766in}{1.080603in}}%
\pgfpathlineto{\pgfqpoint{9.499118in}{1.041124in}}%
\pgfusepath{stroke}%
\end{pgfscope}%
\begin{pgfscope}%
\pgfpathrectangle{\pgfqpoint{0.960541in}{0.793563in}}{\pgfqpoint{8.536910in}{3.807222in}}%
\pgfusepath{clip}%
\pgfsetrectcap%
\pgfsetroundjoin%
\pgfsetlinewidth{3.011250pt}%
\definecolor{currentstroke}{rgb}{0.501961,0.000000,0.501961}%
\pgfsetstrokecolor{currentstroke}%
\pgfsetdash{}{0pt}%
\pgfpathmoveto{\pgfqpoint{0.960541in}{4.413411in}}%
\pgfpathlineto{\pgfqpoint{1.045910in}{4.427729in}}%
\pgfpathlineto{\pgfqpoint{1.131279in}{4.356286in}}%
\pgfpathlineto{\pgfqpoint{1.216649in}{4.359915in}}%
\pgfpathlineto{\pgfqpoint{1.302018in}{4.345597in}}%
\pgfpathlineto{\pgfqpoint{1.336677in}{2.723923in}}%
\pgfpathlineto{\pgfqpoint{1.609859in}{2.709705in}}%
\pgfpathlineto{\pgfqpoint{1.883040in}{2.738640in}}%
\pgfpathlineto{\pgfqpoint{2.156221in}{2.727846in}}%
\pgfpathlineto{\pgfqpoint{2.429402in}{2.740594in}}%
\pgfpathlineto{\pgfqpoint{2.702583in}{2.701489in}}%
\pgfpathlineto{\pgfqpoint{2.975764in}{2.715847in}}%
\pgfpathlineto{\pgfqpoint{3.248945in}{2.707755in}}%
\pgfpathlineto{\pgfqpoint{3.522126in}{2.710158in}}%
\pgfpathlineto{\pgfqpoint{3.795307in}{2.705600in}}%
\pgfpathlineto{\pgfqpoint{4.068489in}{2.715344in}}%
\pgfpathlineto{\pgfqpoint{4.341670in}{2.751872in}}%
\pgfpathlineto{\pgfqpoint{4.614851in}{2.719845in}}%
\pgfpathlineto{\pgfqpoint{4.888032in}{2.735566in}}%
\pgfpathlineto{\pgfqpoint{5.161213in}{2.718706in}}%
\pgfpathlineto{\pgfqpoint{5.434394in}{2.709789in}}%
\pgfpathlineto{\pgfqpoint{5.707575in}{2.718101in}}%
\pgfpathlineto{\pgfqpoint{5.980756in}{2.712302in}}%
\pgfpathlineto{\pgfqpoint{6.253937in}{2.714195in}}%
\pgfpathlineto{\pgfqpoint{6.527119in}{2.702723in}}%
\pgfpathlineto{\pgfqpoint{6.800300in}{2.729047in}}%
\pgfpathlineto{\pgfqpoint{7.073481in}{2.727641in}}%
\pgfpathlineto{\pgfqpoint{7.346662in}{2.725246in}}%
\pgfpathlineto{\pgfqpoint{7.619843in}{2.709630in}}%
\pgfpathlineto{\pgfqpoint{7.893024in}{2.726193in}}%
\pgfpathlineto{\pgfqpoint{8.166205in}{2.686165in}}%
\pgfpathlineto{\pgfqpoint{8.439386in}{2.698624in}}%
\pgfpathlineto{\pgfqpoint{8.712567in}{2.707733in}}%
\pgfpathlineto{\pgfqpoint{8.985749in}{2.716623in}}%
\pgfpathlineto{\pgfqpoint{9.258930in}{1.089000in}}%
\pgfpathlineto{\pgfqpoint{9.326713in}{1.098502in}}%
\pgfpathlineto{\pgfqpoint{9.454766in}{1.080603in}}%
\pgfpathlineto{\pgfqpoint{9.499118in}{1.041124in}}%
\pgfusepath{stroke}%
\end{pgfscope}%
\begin{pgfscope}%
\pgfsetrectcap%
\pgfsetmiterjoin%
\pgfsetlinewidth{2.007500pt}%
\definecolor{currentstroke}{rgb}{0.000000,0.000000,0.000000}%
\pgfsetstrokecolor{currentstroke}%
\pgfsetdash{}{0pt}%
\pgfpathmoveto{\pgfqpoint{0.960541in}{0.793563in}}%
\pgfpathlineto{\pgfqpoint{0.960541in}{4.600785in}}%
\pgfusepath{stroke}%
\end{pgfscope}%
\begin{pgfscope}%
\pgfsetrectcap%
\pgfsetmiterjoin%
\pgfsetlinewidth{2.007500pt}%
\definecolor{currentstroke}{rgb}{0.000000,0.000000,0.000000}%
\pgfsetstrokecolor{currentstroke}%
\pgfsetdash{}{0pt}%
\pgfpathmoveto{\pgfqpoint{9.497451in}{0.793563in}}%
\pgfpathlineto{\pgfqpoint{9.497451in}{4.600785in}}%
\pgfusepath{stroke}%
\end{pgfscope}%
\begin{pgfscope}%
\pgfsetrectcap%
\pgfsetmiterjoin%
\pgfsetlinewidth{2.007500pt}%
\definecolor{currentstroke}{rgb}{0.000000,0.000000,0.000000}%
\pgfsetstrokecolor{currentstroke}%
\pgfsetdash{}{0pt}%
\pgfpathmoveto{\pgfqpoint{0.960541in}{0.793563in}}%
\pgfpathlineto{\pgfqpoint{9.497451in}{0.793563in}}%
\pgfusepath{stroke}%
\end{pgfscope}%
\begin{pgfscope}%
\pgfsetrectcap%
\pgfsetmiterjoin%
\pgfsetlinewidth{2.007500pt}%
\definecolor{currentstroke}{rgb}{0.000000,0.000000,0.000000}%
\pgfsetstrokecolor{currentstroke}%
\pgfsetdash{}{0pt}%
\pgfpathmoveto{\pgfqpoint{0.960541in}{4.600785in}}%
\pgfpathlineto{\pgfqpoint{9.497451in}{4.600785in}}%
\pgfusepath{stroke}%
\end{pgfscope}%
\begin{pgfscope}%
\definecolor{textcolor}{rgb}{0.501961,0.000000,0.000000}%
\pgfsetstrokecolor{textcolor}%
\pgfsetfillcolor{textcolor}%
\pgftext[x=1.218778in, y=2.490903in, left, base,rotate=90.000000]{\color{textcolor}{\sffamily\fontsize{18.000000}{21.600000}\selectfont\catcode`\^=\active\def^{\ifmmode\sp\else\^{}\fi}\catcode`\%=\active\def
\end{pgfscope}%
\begin{pgfscope}%
\definecolor{textcolor}{rgb}{0.000000,0.000000,0.501961}%
\pgfsetstrokecolor{textcolor}%
\pgfsetfillcolor{textcolor}%
\pgftext[x=9.414211in, y=2.490903in, left, base,rotate=90.000000]{\color{textcolor}{\sffamily\fontsize{18.000000}{21.600000}\selectfont\catcode`\^=\active\def^{\ifmmode\sp\else\^{}\fi}\catcode`\%=\active\def
\end{pgfscope}%
\begin{pgfscope}%
\pgfsetrectcap%
\pgfsetroundjoin%
\pgfsetlinewidth{3.011250pt}%
\definecolor{currentstroke}{rgb}{0.000000,0.000000,1.000000}%
\pgfsetstrokecolor{currentstroke}%
\pgfsetdash{}{0pt}%
\pgfpathmoveto{\pgfqpoint{1.622301in}{4.309726in}}%
\pgfpathlineto{\pgfqpoint{1.955635in}{4.309726in}}%
\pgfpathlineto{\pgfqpoint{2.288968in}{4.309726in}}%
\pgfusepath{stroke}%
\end{pgfscope}%
\begin{pgfscope}%
\definecolor{textcolor}{rgb}{0.000000,0.000000,0.000000}%
\pgfsetstrokecolor{textcolor}%
\pgfsetfillcolor{textcolor}%
\pgftext[x=2.466746in,y=4.231948in,left,base]{\color{textcolor}{\sffamily\fontsize{16.000000}{19.200000}\selectfont\catcode`\^=\active\def^{\ifmmode\sp\else\^{}\fi}\catcode`\%=\active\def
\end{pgfscope}%
\begin{pgfscope}%
\pgfsetrectcap%
\pgfsetroundjoin%
\pgfsetlinewidth{3.011250pt}%
\definecolor{currentstroke}{rgb}{0.000000,0.501961,0.000000}%
\pgfsetstrokecolor{currentstroke}%
\pgfsetdash{}{0pt}%
\pgfpathmoveto{\pgfqpoint{3.402944in}{4.309726in}}%
\pgfpathlineto{\pgfqpoint{3.736277in}{4.309726in}}%
\pgfpathlineto{\pgfqpoint{4.069610in}{4.309726in}}%
\pgfusepath{stroke}%
\end{pgfscope}%
\begin{pgfscope}%
\definecolor{textcolor}{rgb}{0.000000,0.000000,0.000000}%
\pgfsetstrokecolor{textcolor}%
\pgfsetfillcolor{textcolor}%
\pgftext[x=4.247388in,y=4.231948in,left,base]{\color{textcolor}{\sffamily\fontsize{16.000000}{19.200000}\selectfont\catcode`\^=\active\def^{\ifmmode\sp\else\^{}\fi}\catcode`\%=\active\def
\end{pgfscope}%
\begin{pgfscope}%
\pgfsetrectcap%
\pgfsetroundjoin%
\pgfsetlinewidth{3.011250pt}%
\definecolor{currentstroke}{rgb}{1.000000,0.647059,0.000000}%
\pgfsetstrokecolor{currentstroke}%
\pgfsetdash{}{0pt}%
\pgfpathmoveto{\pgfqpoint{5.324970in}{4.309726in}}%
\pgfpathlineto{\pgfqpoint{5.658304in}{4.309726in}}%
\pgfpathlineto{\pgfqpoint{5.991637in}{4.309726in}}%
\pgfusepath{stroke}%
\end{pgfscope}%
\begin{pgfscope}%
\definecolor{textcolor}{rgb}{0.000000,0.000000,0.000000}%
\pgfsetstrokecolor{textcolor}%
\pgfsetfillcolor{textcolor}%
\pgftext[x=6.169415in,y=4.231948in,left,base]{\color{textcolor}{\sffamily\fontsize{16.000000}{19.200000}\selectfont\catcode`\^=\active\def^{\ifmmode\sp\else\^{}\fi}\catcode`\%=\active\def
\end{pgfscope}%
\begin{pgfscope}%
\pgfsetrectcap%
\pgfsetroundjoin%
\pgfsetlinewidth{3.011250pt}%
\definecolor{currentstroke}{rgb}{0.501961,0.000000,0.501961}%
\pgfsetstrokecolor{currentstroke}%
\pgfsetdash{}{0pt}%
\pgfpathmoveto{\pgfqpoint{7.246997in}{4.309726in}}%
\pgfpathlineto{\pgfqpoint{7.580331in}{4.309726in}}%
\pgfpathlineto{\pgfqpoint{7.913664in}{4.309726in}}%
\pgfusepath{stroke}%
\end{pgfscope}%
\begin{pgfscope}%
\definecolor{textcolor}{rgb}{0.000000,0.000000,0.000000}%
\pgfsetstrokecolor{textcolor}%
\pgfsetfillcolor{textcolor}%
\pgftext[x=8.091442in,y=4.231948in,left,base]{\color{textcolor}{\sffamily\fontsize{16.000000}{19.200000}\selectfont\catcode`\^=\active\def^{\ifmmode\sp\else\^{}\fi}\catcode`\%=\active\def
\end{pgfscope}%
\begin{pgfscope}%
\pgfpathrectangle{\pgfqpoint{1.463067in}{0.970785in}}{\pgfqpoint{7.532726in}{1.350000in}}%
\pgfusepath{clip}%
\pgfsys@transformshift{1.463067in}{0.970785in}%
\pgftext[left,bottom]{\includegraphics[interpolate=true,width=7.533333in,height=1.350000in]{figures/thermal_profiles-img0.png}}%
\end{pgfscope}%
\end{pgfpicture}%
\makeatother%
\endgroup%

%% file: figures/flux.pgf
\begingroup%
\makeatletter%
\begin{pgfpicture}%
\pgfpathrectangle{\pgfpointorigin}{\pgfqpoint{9.654702in}{4.659755in}}%
\pgfusepath{use as bounding box, clip}%
\begin{pgfscope}%
\pgfsetbuttcap%
\pgfsetmiterjoin%
\definecolor{currentfill}{rgb}{1.000000,1.000000,1.000000}%
\pgfsetfillcolor{currentfill}%
\pgfsetlinewidth{0.000000pt}%
\definecolor{currentstroke}{rgb}{1.000000,1.000000,1.000000}%
\pgfsetstrokecolor{currentstroke}%
\pgfsetdash{}{0pt}%
\pgfpathmoveto{\pgfqpoint{0.000000in}{0.000000in}}%
\pgfpathlineto{\pgfqpoint{9.654702in}{0.000000in}}%
\pgfpathlineto{\pgfqpoint{9.654702in}{4.659755in}}%
\pgfpathlineto{\pgfqpoint{0.000000in}{4.659755in}}%
\pgfpathlineto{\pgfqpoint{0.000000in}{0.000000in}}%
\pgfpathclose%
\pgfusepath{fill}%
\end{pgfscope}%
\begin{pgfscope}%
\pgfsetbuttcap%
\pgfsetmiterjoin%
\definecolor{currentfill}{rgb}{1.000000,1.000000,1.000000}%
\pgfsetfillcolor{currentfill}%
\pgfsetlinewidth{0.000000pt}%
\definecolor{currentstroke}{rgb}{0.000000,0.000000,0.000000}%
\pgfsetstrokecolor{currentstroke}%
\pgfsetstrokeopacity{0.000000}%
\pgfsetdash{}{0pt}%
\pgfpathmoveto{\pgfqpoint{1.054251in}{0.751448in}}%
\pgfpathlineto{\pgfqpoint{9.554702in}{0.751448in}}%
\pgfpathlineto{\pgfqpoint{9.554702in}{4.475337in}}%
\pgfpathlineto{\pgfqpoint{1.054251in}{4.475337in}}%
\pgfpathlineto{\pgfqpoint{1.054251in}{0.751448in}}%
\pgfpathclose%
\pgfusepath{fill}%
\end{pgfscope}%
\begin{pgfscope}%
\pgfpathrectangle{\pgfqpoint{1.054251in}{0.751448in}}{\pgfqpoint{8.500451in}{3.723889in}}%
\pgfusepath{clip}%
\pgfsetbuttcap%
\pgfsetroundjoin%
\definecolor{currentfill}{rgb}{0.000000,0.000000,1.000000}%
\pgfsetfillcolor{currentfill}%
\pgfsetlinewidth{1.003750pt}%
\definecolor{currentstroke}{rgb}{0.000000,0.000000,1.000000}%
\pgfsetstrokecolor{currentstroke}%
\pgfsetdash{}{0pt}%
\pgfsys@defobject{currentmarker}{\pgfqpoint{-0.062113in}{-0.062113in}}{\pgfqpoint{0.062113in}{0.062113in}}{%
\pgfpathmoveto{\pgfqpoint{0.000000in}{-0.062113in}}%
\pgfpathcurveto{\pgfqpoint{0.016473in}{-0.062113in}}{\pgfqpoint{0.032273in}{-0.055568in}}{\pgfqpoint{0.043921in}{-0.043921in}}%
\pgfpathcurveto{\pgfqpoint{0.055568in}{-0.032273in}}{\pgfqpoint{0.062113in}{-0.016473in}}{\pgfqpoint{0.062113in}{0.000000in}}%
\pgfpathcurveto{\pgfqpoint{0.062113in}{0.016473in}}{\pgfqpoint{0.055568in}{0.032273in}}{\pgfqpoint{0.043921in}{0.043921in}}%
\pgfpathcurveto{\pgfqpoint{0.032273in}{0.055568in}}{\pgfqpoint{0.016473in}{0.062113in}}{\pgfqpoint{0.000000in}{0.062113in}}%
\pgfpathcurveto{\pgfqpoint{-0.016473in}{0.062113in}}{\pgfqpoint{-0.032273in}{0.055568in}}{\pgfqpoint{-0.043921in}{0.043921in}}%
\pgfpathcurveto{\pgfqpoint{-0.055568in}{0.032273in}}{\pgfqpoint{-0.062113in}{0.016473in}}{\pgfqpoint{-0.062113in}{0.000000in}}%
\pgfpathcurveto{\pgfqpoint{-0.062113in}{-0.016473in}}{\pgfqpoint{-0.055568in}{-0.032273in}}{\pgfqpoint{-0.043921in}{-0.043921in}}%
\pgfpathcurveto{\pgfqpoint{-0.032273in}{-0.055568in}}{\pgfqpoint{-0.016473in}{-0.062113in}}{\pgfqpoint{0.000000in}{-0.062113in}}%
\pgfpathlineto{\pgfqpoint{0.000000in}{-0.062113in}}%
\pgfpathclose%
\pgfusepath{stroke,fill}%
}%
\begin{pgfscope}%
\pgfsys@transformshift{1.175686in}{1.613454in}%
\pgfsys@useobject{currentmarker}{}%
\end{pgfscope}%
\begin{pgfscope}%
\pgfsys@transformshift{2.997211in}{2.872724in}%
\pgfsys@useobject{currentmarker}{}%
\end{pgfscope}%
\begin{pgfscope}%
\pgfsys@transformshift{4.940171in}{3.711381in}%
\pgfsys@useobject{currentmarker}{}%
\end{pgfscope}%
\begin{pgfscope}%
\pgfsys@transformshift{8.826092in}{3.838366in}%
\pgfsys@useobject{currentmarker}{}%
\end{pgfscope}%
\end{pgfscope}%
\begin{pgfscope}%
\pgfsetbuttcap%
\pgfsetroundjoin%
\definecolor{currentfill}{rgb}{0.000000,0.000000,0.000000}%
\pgfsetfillcolor{currentfill}%
\pgfsetlinewidth{2.007500pt}%
\definecolor{currentstroke}{rgb}{0.000000,0.000000,0.000000}%
\pgfsetstrokecolor{currentstroke}%
\pgfsetdash{}{0pt}%
\pgfsys@defobject{currentmarker}{\pgfqpoint{0.000000in}{0.000000in}}{\pgfqpoint{0.000000in}{0.083333in}}{%
\pgfpathmoveto{\pgfqpoint{0.000000in}{0.000000in}}%
\pgfpathlineto{\pgfqpoint{0.000000in}{0.083333in}}%
\pgfusepath{stroke,fill}%
}%
\begin{pgfscope}%
\pgfsys@transformshift{1.175686in}{0.751448in}%
\pgfsys@useobject{currentmarker}{}%
\end{pgfscope}%
\end{pgfscope}%
\begin{pgfscope}%
\definecolor{textcolor}{rgb}{0.000000,0.000000,0.000000}%
\pgfsetstrokecolor{textcolor}%
\pgfsetfillcolor{textcolor}%
\pgftext[x=1.175686in,y=0.695893in,,top]{\color{textcolor}{\sffamily\fontsize{16.000000}{19.200000}\selectfont\catcode`\^=\active\def^{\ifmmode\sp\else\^{}\fi}\catcode`\%=\active\def
\end{pgfscope}%
\begin{pgfscope}%
\pgfsetbuttcap%
\pgfsetroundjoin%
\definecolor{currentfill}{rgb}{0.000000,0.000000,0.000000}%
\pgfsetfillcolor{currentfill}%
\pgfsetlinewidth{2.007500pt}%
\definecolor{currentstroke}{rgb}{0.000000,0.000000,0.000000}%
\pgfsetstrokecolor{currentstroke}%
\pgfsetdash{}{0pt}%
\pgfsys@defobject{currentmarker}{\pgfqpoint{0.000000in}{0.000000in}}{\pgfqpoint{0.000000in}{0.083333in}}{%
\pgfpathmoveto{\pgfqpoint{0.000000in}{0.000000in}}%
\pgfpathlineto{\pgfqpoint{0.000000in}{0.083333in}}%
\pgfusepath{stroke,fill}%
}%
\begin{pgfscope}%
\pgfsys@transformshift{2.025731in}{0.751448in}%
\pgfsys@useobject{currentmarker}{}%
\end{pgfscope}%
\end{pgfscope}%
\begin{pgfscope}%
\definecolor{textcolor}{rgb}{0.000000,0.000000,0.000000}%
\pgfsetstrokecolor{textcolor}%
\pgfsetfillcolor{textcolor}%
\pgftext[x=2.025731in,y=0.695893in,,top]{\color{textcolor}{\sffamily\fontsize{16.000000}{19.200000}\selectfont\catcode`\^=\active\def^{\ifmmode\sp\else\^{}\fi}\catcode`\%=\active\def
\end{pgfscope}%
\begin{pgfscope}%
\pgfsetbuttcap%
\pgfsetroundjoin%
\definecolor{currentfill}{rgb}{0.000000,0.000000,0.000000}%
\pgfsetfillcolor{currentfill}%
\pgfsetlinewidth{2.007500pt}%
\definecolor{currentstroke}{rgb}{0.000000,0.000000,0.000000}%
\pgfsetstrokecolor{currentstroke}%
\pgfsetdash{}{0pt}%
\pgfsys@defobject{currentmarker}{\pgfqpoint{0.000000in}{0.000000in}}{\pgfqpoint{0.000000in}{0.083333in}}{%
\pgfpathmoveto{\pgfqpoint{0.000000in}{0.000000in}}%
\pgfpathlineto{\pgfqpoint{0.000000in}{0.083333in}}%
\pgfusepath{stroke,fill}%
}%
\begin{pgfscope}%
\pgfsys@transformshift{2.997211in}{0.751448in}%
\pgfsys@useobject{currentmarker}{}%
\end{pgfscope}%
\end{pgfscope}%
\begin{pgfscope}%
\definecolor{textcolor}{rgb}{0.000000,0.000000,0.000000}%
\pgfsetstrokecolor{textcolor}%
\pgfsetfillcolor{textcolor}%
\pgftext[x=2.997211in,y=0.695893in,,top]{\color{textcolor}{\sffamily\fontsize{16.000000}{19.200000}\selectfont\catcode`\^=\active\def^{\ifmmode\sp\else\^{}\fi}\catcode`\%=\active\def
\end{pgfscope}%
\begin{pgfscope}%
\pgfsetbuttcap%
\pgfsetroundjoin%
\definecolor{currentfill}{rgb}{0.000000,0.000000,0.000000}%
\pgfsetfillcolor{currentfill}%
\pgfsetlinewidth{2.007500pt}%
\definecolor{currentstroke}{rgb}{0.000000,0.000000,0.000000}%
\pgfsetstrokecolor{currentstroke}%
\pgfsetdash{}{0pt}%
\pgfsys@defobject{currentmarker}{\pgfqpoint{0.000000in}{0.000000in}}{\pgfqpoint{0.000000in}{0.083333in}}{%
\pgfpathmoveto{\pgfqpoint{0.000000in}{0.000000in}}%
\pgfpathlineto{\pgfqpoint{0.000000in}{0.083333in}}%
\pgfusepath{stroke,fill}%
}%
\begin{pgfscope}%
\pgfsys@transformshift{4.940171in}{0.751448in}%
\pgfsys@useobject{currentmarker}{}%
\end{pgfscope}%
\end{pgfscope}%
\begin{pgfscope}%
\definecolor{textcolor}{rgb}{0.000000,0.000000,0.000000}%
\pgfsetstrokecolor{textcolor}%
\pgfsetfillcolor{textcolor}%
\pgftext[x=4.940171in,y=0.695893in,,top]{\color{textcolor}{\sffamily\fontsize{16.000000}{19.200000}\selectfont\catcode`\^=\active\def^{\ifmmode\sp\else\^{}\fi}\catcode`\%=\active\def
\end{pgfscope}%
\begin{pgfscope}%
\pgfsetbuttcap%
\pgfsetroundjoin%
\definecolor{currentfill}{rgb}{0.000000,0.000000,0.000000}%
\pgfsetfillcolor{currentfill}%
\pgfsetlinewidth{2.007500pt}%
\definecolor{currentstroke}{rgb}{0.000000,0.000000,0.000000}%
\pgfsetstrokecolor{currentstroke}%
\pgfsetdash{}{0pt}%
\pgfsys@defobject{currentmarker}{\pgfqpoint{0.000000in}{0.000000in}}{\pgfqpoint{0.000000in}{0.083333in}}{%
\pgfpathmoveto{\pgfqpoint{0.000000in}{0.000000in}}%
\pgfpathlineto{\pgfqpoint{0.000000in}{0.083333in}}%
\pgfusepath{stroke,fill}%
}%
\begin{pgfscope}%
\pgfsys@transformshift{8.826092in}{0.751448in}%
\pgfsys@useobject{currentmarker}{}%
\end{pgfscope}%
\end{pgfscope}%
\begin{pgfscope}%
\definecolor{textcolor}{rgb}{0.000000,0.000000,0.000000}%
\pgfsetstrokecolor{textcolor}%
\pgfsetfillcolor{textcolor}%
\pgftext[x=8.826092in,y=0.695893in,,top]{\color{textcolor}{\sffamily\fontsize{16.000000}{19.200000}\selectfont\catcode`\^=\active\def^{\ifmmode\sp\else\^{}\fi}\catcode`\%=\active\def
\end{pgfscope}%
\begin{pgfscope}%
\definecolor{textcolor}{rgb}{0.000000,0.000000,0.000000}%
\pgfsetstrokecolor{textcolor}%
\pgfsetfillcolor{textcolor}%
\pgftext[x=5.304476in,y=0.341943in,,top]{\color{textcolor}{\sffamily\fontsize{18.000000}{21.600000}\selectfont\catcode`\^=\active\def^{\ifmmode\sp\else\^{}\fi}\catcode`\%=\active\def
\end{pgfscope}%
\begin{pgfscope}%
\pgfsetbuttcap%
\pgfsetroundjoin%
\definecolor{currentfill}{rgb}{0.000000,0.000000,0.000000}%
\pgfsetfillcolor{currentfill}%
\pgfsetlinewidth{2.007500pt}%
\definecolor{currentstroke}{rgb}{0.000000,0.000000,0.000000}%
\pgfsetstrokecolor{currentstroke}%
\pgfsetdash{}{0pt}%
\pgfsys@defobject{currentmarker}{\pgfqpoint{0.000000in}{0.000000in}}{\pgfqpoint{0.083333in}{0.000000in}}{%
\pgfpathmoveto{\pgfqpoint{0.000000in}{0.000000in}}%
\pgfpathlineto{\pgfqpoint{0.083333in}{0.000000in}}%
\pgfusepath{stroke,fill}%
}%
\begin{pgfscope}%
\pgfsys@transformshift{1.054251in}{0.751448in}%
\pgfsys@useobject{currentmarker}{}%
\end{pgfscope}%
\end{pgfscope}%
\begin{pgfscope}%
\definecolor{textcolor}{rgb}{0.000000,0.000000,0.000000}%
\pgfsetstrokecolor{textcolor}%
\pgfsetfillcolor{textcolor}%
\pgftext[x=0.503903in, y=0.667030in, left, base]{\color{textcolor}{\sffamily\fontsize{16.000000}{19.200000}\selectfont\catcode`\^=\active\def^{\ifmmode\sp\else\^{}\fi}\catcode`\%=\active\def
\end{pgfscope}%
\begin{pgfscope}%
\pgfsetbuttcap%
\pgfsetroundjoin%
\definecolor{currentfill}{rgb}{0.000000,0.000000,0.000000}%
\pgfsetfillcolor{currentfill}%
\pgfsetlinewidth{2.007500pt}%
\definecolor{currentstroke}{rgb}{0.000000,0.000000,0.000000}%
\pgfsetstrokecolor{currentstroke}%
\pgfsetdash{}{0pt}%
\pgfsys@defobject{currentmarker}{\pgfqpoint{0.000000in}{0.000000in}}{\pgfqpoint{0.083333in}{0.000000in}}{%
\pgfpathmoveto{\pgfqpoint{0.000000in}{0.000000in}}%
\pgfpathlineto{\pgfqpoint{0.083333in}{0.000000in}}%
\pgfusepath{stroke,fill}%
}%
\begin{pgfscope}%
\pgfsys@transformshift{1.054251in}{1.496226in}%
\pgfsys@useobject{currentmarker}{}%
\end{pgfscope}%
\end{pgfscope}%
\begin{pgfscope}%
\definecolor{textcolor}{rgb}{0.000000,0.000000,0.000000}%
\pgfsetstrokecolor{textcolor}%
\pgfsetfillcolor{textcolor}%
\pgftext[x=0.503903in, y=1.411808in, left, base]{\color{textcolor}{\sffamily\fontsize{16.000000}{19.200000}\selectfont\catcode`\^=\active\def^{\ifmmode\sp\else\^{}\fi}\catcode`\%=\active\def
\end{pgfscope}%
\begin{pgfscope}%
\pgfsetbuttcap%
\pgfsetroundjoin%
\definecolor{currentfill}{rgb}{0.000000,0.000000,0.000000}%
\pgfsetfillcolor{currentfill}%
\pgfsetlinewidth{2.007500pt}%
\definecolor{currentstroke}{rgb}{0.000000,0.000000,0.000000}%
\pgfsetstrokecolor{currentstroke}%
\pgfsetdash{}{0pt}%
\pgfsys@defobject{currentmarker}{\pgfqpoint{0.000000in}{0.000000in}}{\pgfqpoint{0.083333in}{0.000000in}}{%
\pgfpathmoveto{\pgfqpoint{0.000000in}{0.000000in}}%
\pgfpathlineto{\pgfqpoint{0.083333in}{0.000000in}}%
\pgfusepath{stroke,fill}%
}%
\begin{pgfscope}%
\pgfsys@transformshift{1.054251in}{2.241004in}%
\pgfsys@useobject{currentmarker}{}%
\end{pgfscope}%
\end{pgfscope}%
\begin{pgfscope}%
\definecolor{textcolor}{rgb}{0.000000,0.000000,0.000000}%
\pgfsetstrokecolor{textcolor}%
\pgfsetfillcolor{textcolor}%
\pgftext[x=0.503903in, y=2.156585in, left, base]{\color{textcolor}{\sffamily\fontsize{16.000000}{19.200000}\selectfont\catcode`\^=\active\def^{\ifmmode\sp\else\^{}\fi}\catcode`\%=\active\def
\end{pgfscope}%
\begin{pgfscope}%
\pgfsetbuttcap%
\pgfsetroundjoin%
\definecolor{currentfill}{rgb}{0.000000,0.000000,0.000000}%
\pgfsetfillcolor{currentfill}%
\pgfsetlinewidth{2.007500pt}%
\definecolor{currentstroke}{rgb}{0.000000,0.000000,0.000000}%
\pgfsetstrokecolor{currentstroke}%
\pgfsetdash{}{0pt}%
\pgfsys@defobject{currentmarker}{\pgfqpoint{0.000000in}{0.000000in}}{\pgfqpoint{0.083333in}{0.000000in}}{%
\pgfpathmoveto{\pgfqpoint{0.000000in}{0.000000in}}%
\pgfpathlineto{\pgfqpoint{0.083333in}{0.000000in}}%
\pgfusepath{stroke,fill}%
}%
\begin{pgfscope}%
\pgfsys@transformshift{1.054251in}{2.985782in}%
\pgfsys@useobject{currentmarker}{}%
\end{pgfscope}%
\end{pgfscope}%
\begin{pgfscope}%
\definecolor{textcolor}{rgb}{0.000000,0.000000,0.000000}%
\pgfsetstrokecolor{textcolor}%
\pgfsetfillcolor{textcolor}%
\pgftext[x=0.503903in, y=2.901363in, left, base]{\color{textcolor}{\sffamily\fontsize{16.000000}{19.200000}\selectfont\catcode`\^=\active\def^{\ifmmode\sp\else\^{}\fi}\catcode`\%=\active\def
\end{pgfscope}%
\begin{pgfscope}%
\pgfsetbuttcap%
\pgfsetroundjoin%
\definecolor{currentfill}{rgb}{0.000000,0.000000,0.000000}%
\pgfsetfillcolor{currentfill}%
\pgfsetlinewidth{2.007500pt}%
\definecolor{currentstroke}{rgb}{0.000000,0.000000,0.000000}%
\pgfsetstrokecolor{currentstroke}%
\pgfsetdash{}{0pt}%
\pgfsys@defobject{currentmarker}{\pgfqpoint{0.000000in}{0.000000in}}{\pgfqpoint{0.083333in}{0.000000in}}{%
\pgfpathmoveto{\pgfqpoint{0.000000in}{0.000000in}}%
\pgfpathlineto{\pgfqpoint{0.083333in}{0.000000in}}%
\pgfusepath{stroke,fill}%
}%
\begin{pgfscope}%
\pgfsys@transformshift{1.054251in}{3.730559in}%
\pgfsys@useobject{currentmarker}{}%
\end{pgfscope}%
\end{pgfscope}%
\begin{pgfscope}%
\definecolor{textcolor}{rgb}{0.000000,0.000000,0.000000}%
\pgfsetstrokecolor{textcolor}%
\pgfsetfillcolor{textcolor}%
\pgftext[x=0.503903in, y=3.646141in, left, base]{\color{textcolor}{\sffamily\fontsize{16.000000}{19.200000}\selectfont\catcode`\^=\active\def^{\ifmmode\sp\else\^{}\fi}\catcode`\%=\active\def
\end{pgfscope}%
\begin{pgfscope}%
\pgfsetbuttcap%
\pgfsetroundjoin%
\definecolor{currentfill}{rgb}{0.000000,0.000000,0.000000}%
\pgfsetfillcolor{currentfill}%
\pgfsetlinewidth{2.007500pt}%
\definecolor{currentstroke}{rgb}{0.000000,0.000000,0.000000}%
\pgfsetstrokecolor{currentstroke}%
\pgfsetdash{}{0pt}%
\pgfsys@defobject{currentmarker}{\pgfqpoint{0.000000in}{0.000000in}}{\pgfqpoint{0.083333in}{0.000000in}}{%
\pgfpathmoveto{\pgfqpoint{0.000000in}{0.000000in}}%
\pgfpathlineto{\pgfqpoint{0.083333in}{0.000000in}}%
\pgfusepath{stroke,fill}%
}%
\begin{pgfscope}%
\pgfsys@transformshift{1.054251in}{4.475337in}%
\pgfsys@useobject{currentmarker}{}%
\end{pgfscope}%
\end{pgfscope}%
\begin{pgfscope}%
\definecolor{textcolor}{rgb}{0.000000,0.000000,0.000000}%
\pgfsetstrokecolor{textcolor}%
\pgfsetfillcolor{textcolor}%
\pgftext[x=0.503903in, y=4.390919in, left, base]{\color{textcolor}{\sffamily\fontsize{16.000000}{19.200000}\selectfont\catcode`\^=\active\def^{\ifmmode\sp\else\^{}\fi}\catcode`\%=\active\def
\end{pgfscope}%
\begin{pgfscope}%
\definecolor{textcolor}{rgb}{0.000000,0.000000,0.000000}%
\pgfsetstrokecolor{textcolor}%
\pgfsetfillcolor{textcolor}%
\pgftext[x=0.365014in,y=2.613393in,,bottom,rotate=90.000000]{\color{textcolor}{\sffamily\fontsize{18.000000}{21.600000}\selectfont\catcode`\^=\active\def^{\ifmmode\sp\else\^{}\fi}\catcode`\%=\active\def
\end{pgfscope}%
\begin{pgfscope}%
\pgfpathrectangle{\pgfqpoint{1.054251in}{0.751448in}}{\pgfqpoint{8.500451in}{3.723889in}}%
\pgfusepath{clip}%
\pgfsetrectcap%
\pgfsetroundjoin%
\pgfsetlinewidth{2.509375pt}%
\definecolor{currentstroke}{rgb}{0.000000,0.000000,1.000000}%
\pgfsetstrokecolor{currentstroke}%
\pgfsetdash{}{0pt}%
\pgfpathmoveto{\pgfqpoint{1.175686in}{1.613454in}}%
\pgfpathlineto{\pgfqpoint{2.997211in}{2.872724in}}%
\pgfpathlineto{\pgfqpoint{4.940171in}{3.711381in}}%
\pgfpathlineto{\pgfqpoint{8.826092in}{3.838366in}}%
\pgfusepath{stroke}%
\end{pgfscope}%
\begin{pgfscope}%
\pgfsetrectcap%
\pgfsetmiterjoin%
\pgfsetlinewidth{2.007500pt}%
\definecolor{currentstroke}{rgb}{0.000000,0.000000,0.000000}%
\pgfsetstrokecolor{currentstroke}%
\pgfsetdash{}{0pt}%
\pgfpathmoveto{\pgfqpoint{1.054251in}{0.751448in}}%
\pgfpathlineto{\pgfqpoint{1.054251in}{4.475337in}}%
\pgfusepath{stroke}%
\end{pgfscope}%
\begin{pgfscope}%
\pgfsetrectcap%
\pgfsetmiterjoin%
\pgfsetlinewidth{2.007500pt}%
\definecolor{currentstroke}{rgb}{0.000000,0.000000,0.000000}%
\pgfsetstrokecolor{currentstroke}%
\pgfsetdash{}{0pt}%
\pgfpathmoveto{\pgfqpoint{9.554702in}{0.751448in}}%
\pgfpathlineto{\pgfqpoint{9.554702in}{4.475337in}}%
\pgfusepath{stroke}%
\end{pgfscope}%
\begin{pgfscope}%
\pgfsetrectcap%
\pgfsetmiterjoin%
\pgfsetlinewidth{2.007500pt}%
\definecolor{currentstroke}{rgb}{0.000000,0.000000,0.000000}%
\pgfsetstrokecolor{currentstroke}%
\pgfsetdash{}{0pt}%
\pgfpathmoveto{\pgfqpoint{1.054251in}{0.751448in}}%
\pgfpathlineto{\pgfqpoint{9.554702in}{0.751448in}}%
\pgfusepath{stroke}%
\end{pgfscope}%
\begin{pgfscope}%
\pgfsetrectcap%
\pgfsetmiterjoin%
\pgfsetlinewidth{2.007500pt}%
\definecolor{currentstroke}{rgb}{0.000000,0.000000,0.000000}%
\pgfsetstrokecolor{currentstroke}%
\pgfsetdash{}{0pt}%
\pgfpathmoveto{\pgfqpoint{1.054251in}{4.475337in}}%
\pgfpathlineto{\pgfqpoint{9.554702in}{4.475337in}}%
\pgfusepath{stroke}%
\end{pgfscope}%
\end{pgfpicture}%
\makeatother%
\endgroup%